\documentclass[iop]{emulateapj}

\UseRawInputEncoding 
\usepackage{amsmath,natbib,graphicx}
\usepackage{epsf}
\usepackage{epstopdf}
\usepackage{epsfig}
\usepackage{color}
\DeclareGraphicsExtensions{.jpg,.pdf,.png,.eps,.ps}
\usepackage{scrextend}
\usepackage{hyperref}

\newcommand{\II}{~{\sc ii}}
\newcommand{\III}{~{\sc iii}}

\shorttitle{}

\begin{document}

\title{Extended hard X-ray emission in highly obscured AGN}

\author{Jingzhe Ma$^{1}$, Martin Elvis$^{1}$, G. Fabbiano$^{1}$, Mislav Balokovi{\'c}$^{2,3}$, W. Peter Maksym$^{1}$, and Guido Risaliti$^{4}$ }

\altaffiltext{1}{Center for Astrophysics $|$ Harvard \& Smithsonian, 60 Garden St, Cambridge, MA 02138, USA; \href{jingzhe.ma@cfa.harvard.edu}{jingzhe.ma@cfa.harvard.edu}}
\altaffiltext{2}{Yale Center for Astronomy \& Astrophysics, 52 Hillhouse Avenue, New Haven, CT 06511, USA}
\altaffiltext{3}{Department of Physics, Yale University, P.O. Box 208120, New Haven, CT 06520, USA}
\altaffiltext{4}{INAF- Osservatorio Astrofisico di Arcetri, Largo E. Fermi 5 50125 Firenze, Italy}


\begin{abstract}

Kilo-parsec scale hard ($>$ 3 keV) X-ray continuum and fluorescent Fe K$\alpha$ line emission has been recently discovered in nearby Compton-thick (CT) active galactic nuclei (AGN), which opens new opportunities to improve AGN torus modeling and investigate how the central supermassive black hole interacts with and impacts the host galaxy. Following a pilot Chandra survey of nearby CT AGN, we present in this paper the Chandra spatial analysis results of five uniformly selected non-CT but still heavily obscured AGN to investigate the extended hard X-ray emission by measuring the excess emission counts, excess fractions, and physical scales. Three of them show extended emission in the 3.0-7.0 keV band detected at $>$ 3$\sigma$ above the Chandra PSF with total excess fractions ranging from $\sim$8\% - 20\%. The extent of the hard emission ranges from at least $\sim$250 pc to 1.1 kpc in radius. We compare these new sources with CT AGN and find that CT AGN appear to be more extended in the hard band than the non-CT AGN. Similar to CT AGN, the amounts of extended hard X-ray emission relative to the total emission of these obscured AGN are not negligible. Together with other extended hard X-ray detected AGN in the literature, we further explore potential correlations between the extended hard X-ray component and AGN parameters. We also discuss the implications for torus modeling and AGN feedback. Considering potential contributions from X-ray binaries (XRBs) to the extended emission, we do not see strong XRB contamination in the overall sample.

\end{abstract}


\section{Introduction}
\label{sec:intro}

Recently, extended hard ($>$ 3 keV) X-ray emission, both the hard continuum and fluorescent Fe K$\alpha$ lines, has been found in several nearby Compton-thick (CT) active galactic nuclei (AGN) using deep Chandra imaging (e.g., \citealt{Arevalo2014,Bauer2015,Maksym2017,Fabbiano2017,Fabbiano2018a,Fabbiano2018b,Fabbiano2019,Jones2020,Travascio2021}), reaching out from the central supermassive black hole (SMBH) to $\sim$kpc-scales. This emission has consequences for both the feedback of AGN on their host galaxies, and for the properties of the central obscuring torus. The hard X-ray emission extends not only along the direction of the ionization cones but also in the region perpendicular to the ionization cones, i.e., cross-cones (e.g., \citealt{Fabbiano2018a,Fabbiano2018b,Fabbiano2019,Jones2020}). However, in the standard AGN unified model, the characteristic hard X- ray continuum and fluorescent Fe K$\alpha$ lines are confined to the nuclear surroundings as the obscuring torus should completely obscure the nucleus (e.g., \citealt{Antonucci1993, Urry1995,Netzer2015}), complicating the determination of the torus properties.

Rather than a completely obscuring torus in the standard AGN unified model, the appearance of extended hard X-ray emission supports the increasingly popular scenario of a clumpy structure of the torus (e.g., \citealt{Nenkova2008,Elitzur2012}), which allows for the transmission of radiation on kpc scales. 

The interactions of the photons escaping the nuclear region with the interstellar medium (ISM) clouds in the host galaxy would give rise to the extended diffuse emission in both the ionization cone and cross-cone directions. For example, the extended hard continuum and Fe K$\alpha$ emission observed in ESO 428-G014, a well-studied nearby CT AGN, is likely caused by scattering off dense molecular clouds in the host galaxy of photons escaping the nuclear region (e.g., \citealt{Fabbiano2018a,Fabbiano2018b,Fabbiano2019}). The kpc-scale spatial extent is larger at the lower energies, suggesting that the optically thick molecular clouds responsible for the scattering of the higher energy photons are more concentrated in the inner radii.

The previous discoveries were based on individual objects. We initiated a joint Chandra and NuSTAR survey of nearby heavily obscured AGN systematically selected from the {\it Swift}-BAT spectroscopic AGN survey (BASS; \citealt{Koss2017,Ricci2017}) to specifically investigate the extended hard X-ray emission. Our pilot Chandra Cycle 20 survey consists of seven CT AGN selected from the 70-month catalog \citep{Koss2017}. We quantitatively measured the amount and extent of the extended component above the Chandra point-spread-function (PSF). Five out of the seven CT AGN show extended emission in the 3-7 keV band detected at $>$ 3$\sigma$ above the PSF, with $\sim$12\%-22\% of the total emission in the extended components \citep{Ma2020}. ESO 137-G034 and NGC 3281 display biconical ionization structures with extended hard X-ray emission reaching kpc-scales ($\sim$1.9 kpc and 3.5 kpc in diameter). The other three show extended hard X-ray emission above the PSF out to at least $\sim$360 pc in radius. We further explored potential correlations between the measured quantities with physical parameters such as AGN bolometric luminosity, column density $N_{\rm H}$ etc. There appeared to be a moderate correlation between the total extended excess fraction and log$N_{\rm H}$ \citep{Ma2020}. However, a definite correlation could not be drawn due to the small sample size. Given that this extended hard X-ray component appears to be relatively common in this uniformly selected CT AGN sample, we further discussed the implications for torus modeling and AGN feedback. Detecting hard X-ray emission beyond the traditional dusty torus in the standard AGN unified model implies that we need to test and improve torus modeling and update our knowledge of SMBH-host galaxy interactions. 

Following the pilot survey, we have been expanding the survey in an effort to build a well-defined statistical sample in order to answer the following questions: (1) What are the typical ranges of extent and amount of extended hard X-ray emission relative to the total emission? (2) What is the origin of the extended hard X-ray emission? (3) What are the implications on AGN models? (4) What are the implications for AGN feedback?

In this paper, we report the results from our Chandra Cycle 22 imaging program on five heavily obscured AGN but not CT with 23.0 $<$ log$N_{\rm H}$ $<$ 23.9 cm$^{-2}$. We describe the sample selection, Chandra/ACIS-S observations, and data reduction in Section \ref{obs}. Section \ref{methods} introduces the spatial analysis methods we utilized in this work. We present the results for individual AGN in Section \ref{results} and compare with other extended hard X-ray detected AGN in the literature in Section \ref{discussion}. We also discuss implications for torus modeling and AGN feedback. Section \ref{summary} summarizes the conclusions and plans for future observations.

\begin{table*}
\centering
\caption{Observation Log}
\scriptsize
\begin{tabular}{lccccccc}
\hline\hline
Sourcename & $z$   &ObsID    & Instrument    &  $T_{\rm exp}$ (ks)&  PI  & Date                    &  Net Counts \\  
 \hline
NGC 678          &0.00946   &  23808 & ACIS-S          & 37.16                & Elvis &   2020 Sept 29    & 401 $\pm$ 21 \\    
IC 1657             & 0.01195    &  23809   & ACIS-S         & 14.33              & Elvis & 2021 May 19      & 360 $\pm$ 20 \\   
NGC 5899        &  0.00864  & 23810   & ACIS-S              & 9.57               & Elvis    & 2020 Dec 7   & 1080 $\pm$ 33 \\		
NGC 454E        & 0.01213   & 23812  & ACIS-S             & 14.33                & Elvis  & 2021 Feb 14    & 167 $\pm$ 13  \\		
ESO 234-G050 & 0.00877 & 23814  & ACIS-S            & 19.08               & Elvis  & 2021 Mar 14    &  316 $\pm$ 20 \\			
\hline
\end{tabular}
\tablecomments{The redshifts are taken from NED. The total net counts (background subtracted) at 0.3-7.0 keV are listed in the last column. }
\label{table1}
\end{table*}

\section{Sample selection, observations, and data reduction}
\label{obs}

The five heavily obscured AGN are the targets of our joint Chandra Cycle 22 and NuSTAR program (P.I. M. Elvis). We selected these sources from the {\it Swift}-BAT spectroscopic AGN survey 70-month catalog \citep{Koss2017}, based on the following criteria: (1) $z < 0.013$, $D$ $<$ 50 Mpc, which gives a plate scale of 1$\arcsec$ $\sim$ 250 pc, such that the extended emission can be spatially resolved with sub-arcsecond Chandra ACIS-S imaging; (2) 23.0 $<$ log $N_{\rm H}$ $<$ 23.9 cm$^{-2}$ from \cite{Ricci2017} to ensure nuclear obscuration is present; (3) 2-10 keV BASS flux of $>$ 4 $\times$ 10$^{-13}$ erg cm$^{-2}$ s$^{-1}$ to ensure an adequate count rate.

Table \ref{table1} summarizes the Chandra Cycle 22 observations for this work. We first reprocessed the data using CIAO\footnote{CIAO; http://cxc.harvard.edu/ciao/} (v4.13) and CALDB\footnote{CALDB; http://cxc.harvard.edu/caldb/} (v4.9.6) \citep{Fruscione2006}, provided by the Chandra X-ray Center (CXC). The default parameters in chandra\_repro were adopted. We examined the high background flares, and the entire dataset was acceptable. Pile-up is not a concern given the low count rates and the (1/4) sub-array\footnote{We used the central 1/4 sub-array (256 rows starting from Row 385) at the aim point.} configurations of our observations.

\begin{table*}
\centering
\caption{The excess counts over the Chandra PSF, extended fractions, and total excess fractions. }
\begin{tabular}{lccccc}
\hline\hline
Sourcename &  \bf{ 0.3-3.0 keV}                                           &\bf{ 0.3-3.0 keV}                      & \bf{ 0.3-3.0 keV}                                                   &\bf{ 0.3-3.0 keV}    \\
            &     excess counts $\geq$1.5$\arcsec$      & total excess counts            & extended fraction $\geq$1.5$\arcsec$   & total excess fraction    \\
\hline
NGC 678          &  35.6 $\pm$ 7.3 (4.9$\sigma$)    & 61.2 $\pm$ 9.2 (6.6$\sigma$)     &  42.5\% $\pm$ 10.1\%              &     72.1\% $\pm$ 13.9\%       \\
IC 1657           & 42.8 $\pm$ 8.7 (4.9$\sigma$)      & 48.2 $\pm$ 9.2 (5.2$\sigma$)   & 67.4\% $\pm$ 17.2\%               &75.9\% $\pm$ 18.7\%      \\
NGC 5899       &  $<$ 6.9                                        & 17.0 $\pm$ 8.3 (2.1$\sigma$)            &$<$ 5.3\%                        &9.3\% $\pm$ 4.6\%     \\
NGC 454E       &   5.9 $\pm$ 3.7 (1.6$\sigma$)           & 20.9 $\pm$ 6.1 (3.4$\sigma$)         &9.6\% $\pm$ 6.1\%           &34.0\% $\pm$ 11.0\%             \\
ESO 234-G050       &  49.3 $\pm$ 10.4 (4.7$\sigma$)    & 91.3 $\pm$ 12.6 (7.5$\sigma$)       &36.7\% $\pm$ 8.6\%           &68.1\% $\pm$ 11.7\%         \\
\hline

             &     \bf{3.0-7.0 keV}   & \bf{3.0-7.0 keV}                 &     \bf{3.0-7.0 keV}   & \bf{3.0-7.0 keV}  \\
\hline 
NGC 678        &     10.9 $\pm$ 5.5 (2.0$\sigma$)      &49.5 $\pm$ 11.1 (4.4$\sigma$)        & 4.3\% $\pm$ 2.2\%                      & 19.7\% $\pm$ 4.6\% \\
IC 1657       &   10.0 $\pm$ 7.0 (1.4$\sigma$)             &57.7 $\pm$ 12.6 (4.6$\sigma$)                & 3.4\% $\pm$ 2.4\%                      &  19.4\% $\pm$ 4.4\%    \\
NGC 5899   &  $<$ 22.3                                            &72.7 $\pm$ 18.6 (3.9$\sigma$)             & $<$ 2.5\%                                   &8.1\% $\pm$ 2.1\%     \\
NGC 454E         &  $<$ 10.5                                     & 20.3 $\pm$ 7.2 (2.8$\sigma$)         &  $<$ 10.0\%                                &19.2\% $\pm$ 7.1\%  \\
ESO 234-G050       &  12.5 $\pm$ 7.5 (1.7$\sigma$)    & 31.0 $\pm$ 10.7 (2.9$\sigma$)         &  6.9\% $\pm$ 4.2\%                  & 17.1\% $\pm$ 6.1\%   \\
\hline

             &     \bf{6.0-7.0 keV}   & \bf{6.0-7.0 keV}                 &     \bf{6.0-7.0 keV}   & \bf{6.0-7.0 keV}  \\
\hline 
NGC 678  & 5.2 $\pm$ 3.0 (1.7$\sigma$)   &26.9 $\pm$ 6.6 (4.1$\sigma$)     & 7.5\% $\pm$ 4.5\%   &38.7\% $\pm$ 10.6\%\\
\hline
\hline
\end{tabular}
\label{table2}
\tablecomments{Extended fraction $\equiv$ (Source 1.5$\arcsec$-8$\arcsec$ counts $-$ PSF 1.5$\arcsec$-8$\arcsec$ counts) / (Total 8$\arcsec$ source counts). Total excess fraction $\equiv$ (Source 0.5$\arcsec$-8$\arcsec$ counts $-$ PSF 0.5$\arcsec$-8$\arcsec$ counts) / (Total 8$\arcsec$ source counts). We place a 3$\sigma$ upper limit if there are no excess counts above the PSF (i.e. $<$ 1$\sigma$). For IC 1657 and ESO 234-G050, the outer radius is 17$\arcsec$.  }
\end{table*}

\begin{table*}
\centering
\caption{The excess counts over the Chandra PSF in azimuthal sectors.  }
\begin{tabular}{lccc}
\hline\hline
Sourcename & Sector&  \bf{ 0.3-3.0 keV}                                           &\bf{ 0.3-3.0 keV}                     \\
            &  &   excess counts $\geq$1.5$\arcsec$      & total excess counts          \\
\hline
IC 1657           &soft X-ray elongation&37.9 $\pm$ 7.1 (5.3$\sigma$)      & 40.0 $\pm$ 7.4 (5.4$\sigma$)       \\
			& cross sector&4.9 $\pm$ 4.2 (1.2$\sigma$)      & 8.2 $\pm$ 4.7 (1.7$\sigma$)       \\
ESO 234-G050   &soft X-ray elongation    &  46.1 $\pm$ 8.4 (5.5$\sigma$)    & 65.4 $\pm$ 9.7 (6.7$\sigma$)            \\
			& cross sector &$<$ 15.6    & 25.9 $\pm$ 7.3 (3.5$\sigma$)       \\

\hline

             &   &  \bf{3.0-7.0 keV}   & \bf{3.0-7.0 keV}                  \\
           &   &  excess counts $\geq$1.5$\arcsec$         &total excess counts          \\
\hline 
IC 1657   &soft X-ray elongation    &   7.3 $\pm$ 4.9 (1.5$\sigma$)             &24.9 $\pm$ 8.7 (2.9$\sigma$)          \\
                 & cross sector    &  $<$ 13.4             &33.5 $\pm$ 8.8 (3.8$\sigma$)          \\
ESO 234-G050   & soft X-ray elongation    &  10.7 $\pm$ 5.4 (2.0$\sigma$)    & 17.3 $\pm$ 7.6 (2.3$\sigma$)        \\
		& cross sector &$<$ 13.8    & 14.0 $\pm$ 7.1 (2.0$\sigma$)       \\

\hline
\hline
\end{tabular}
\label{table3}
\tablecomments{ We place a 3$\sigma$ upper limit if there are no excess counts above the PSF (i.e. $<$ 1$\sigma$). }
\end{table*}

\section{Spatial analysis methods}
\label{methods}

\subsection{Sub-pixel imaging} 

We investigated the X-ray morphological properties of the obscured AGN using the CIAO image analysis tools installed in SAOImage DS9\footnote{ds9; http://ds9.si.edu}. Images were created in the 0.3-7.0 keV band (full band), the 0.3-3.0 keV band (soft band), the 3.0-7.0 keV band (hard band), and the 6.0-7.0 keV band (where the Fe K$\alpha$ could dominate) in the following analysis. We employed the sub-pixel binning technique to push for higher spatial resolution, which has been tested and frequently applied to imaging studies of extended emission and X-ray jets (e.g., \citealt{Wang2011a,Wang2011b,Wang2011c,Paggi2012}). We used a fine pixel size of 0.062$\arcsec$ (1/8 of the ACIS native pixel size) when producing the images.

\subsection{Radial profiles}
\label{sec3.2}

To quantitatively measure the extent and amount of extended emission, we generated radial surface brightness profiles in different energy bands and different azimuthal sectors (when possible), following the procedure described in our previous work (e.g., \citealt{Fabbiano2017, Fabbiano2018a,Jones2020, Ma2020}). We extracted radial profiles using concentric annuli out to a radius of 8$\arcsec$ for three sources and a radius of 17$\arcsec$ for two sources, reaching the background level. Off-nuclear point sources within the outer circle were all removed before generating the radial profiles. We started with an annular bin size of 0.5$\arcsec$ and increased the bin size at larger radii to maintain a minimum of 10 counts in each bin. To gauge the magnitude and significance of the extended emission, we compared the radial profiles to the Chandra PSFs for the corresponding energy bands. We modeled the PSF for each given centroid position and energy band using ChaRT\footnote{https://cxc.harvard.edu/ciao/PSFs/chart2/} and MARX 5.5.0\footnote{https://space.mit.edu/cxc/marx} following the CIAO PSF simulation thread\footnote{https://cxc.cfa.harvard.edu/ciao/threads/psf.html}. The default AspectBlur parameter in {\it simulate\_psf}\footnote{https://cxc.cfa.harvard.edu/ciao/ahelp/simulate\_psf.html} (blur = 0.07$\arcsec$) was adopted. The radial profiles show the background-subtracted\footnote{We used annular background regions centered around the targets at $\sim$25$\arcsec$(inner) to $\sim$40$\arcsec$(outer) in radius, avoiding regions with point sources.} surface brightness distribution in units of counts per arcsecond$^2$ in each energy band. The PSF radial profiles were generated in the same energy bands and were normalized to the counts within the central 0.5$\arcsec$ radius bin. 

Following \cite{Ma2020}, we take several measures to quantify the amount and extent of the extended emission in each energy band. We first measured the excess counts over the Chandra PSF outside the 1.5$\arcsec$ radius circle to avoid potential contamination from a nuclear component. We also measured the total excess counts above the PSF, which includes all the extended emission that does not belong to the central point source. Along with the excess counts, we also measured the extended, excess fractions in each energy band. The extended fraction $>$ 1.5$\arcsec$ is defined as the ratio of the excess counts above the Chandra PSF in the 1.5$\arcsec$-8$\arcsec$ or 1.5$\arcsec$-17$\arcsec$ annular region to the background-subtracted, total counts within the 8$\arcsec$ or 17$\arcsec$ radius circle at the given energy band. We define the total excess fraction to be the ratio of the total excess counts above the PSF (including the 0.5$\arcsec$-1.5$\arcsec$ region) to the total net counts within the 8$\arcsec$ or 17$\arcsec$ radius circle at the given energy band. Table \ref{table2} lists the excess counts over the Chandra PSF with associated Poisson statistical errors (including the background error), the fraction of the extended emission in the 1.5$\arcsec$-8$\arcsec$ or 1.5$\arcsec$-17$\arcsec$ annular region, and the total excess fraction in each energy band. In cases where we were able to identify azimuthal sectors, e.g., ionization cones and cross-cones, we also measured excess counts in each sector (Table \ref{table3}).  

In addition to the excess counts and excess fractions, we also estimated the fluxes and luminosities of the extended emission (Table \ref{table4}). We converted the 0.3-3.0 keV and 3.0-7.0 keV excess counts to fluxes and luminosities assuming an absorbed (Galactic absorption) power-law model with a photon index $\Gamma$ of 1.9 for all the sources. Variations on the assumed $\Gamma$ from 1.4 to 2.4 change the estimated fluxes/luminosities by less than 20\%. 

The above-mentioned measurements are based on the simulated Chandra PSF. Potential caveats due to the uncertainties of the simulated PSF are discussed in the Appendix.

We discuss the morphology, extended component, and radial profiles of individual sources in the following section.

\section{Results}
\label{results}

\subsection{NGC 678}

NGC 678 is classified as an edge-on, barred spiral galaxy [SB(s)b] at $z$ = 0.00946 (NED; $D$ $\sim$ 42 Mpc; 1$\arcsec$ $\sim$ 200 pc) with a column density of log $N_{\rm H}$ = 23.40 $^{+0.59}_{-0.50}$ cm$^{-2}$ \citep{Ricci2017}.

\begin{figure*}
\centering
\includegraphics[width=13.5cm]{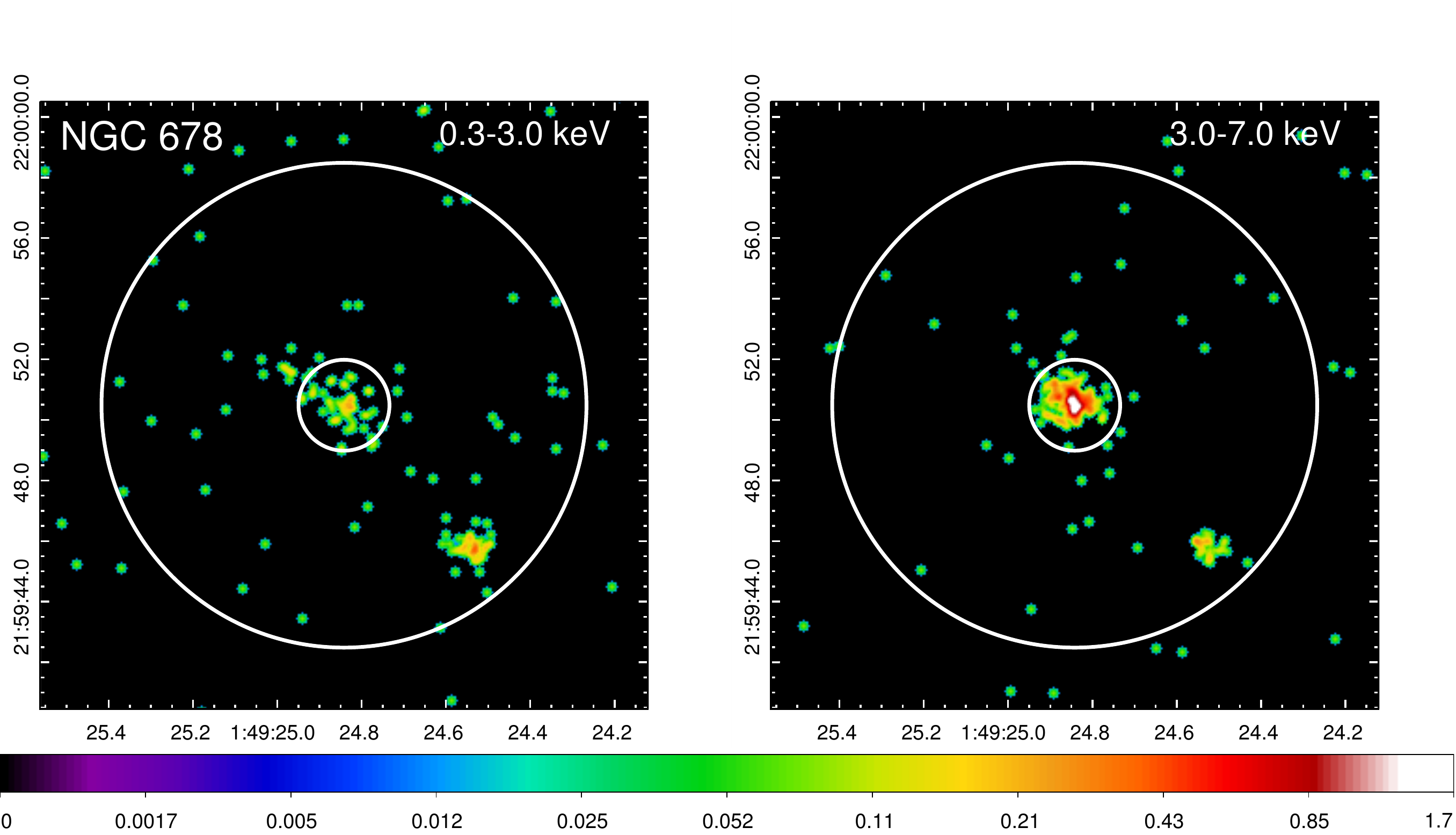} 
\caption{20$\arcsec$ $\times$ 20$\arcsec$ Chandra ACIS-S 0.3-3.0 keV (left) and 3.0-7.0 keV (right) band images of NGC 678 at 1/8 subpixel binning (slightly smoothed with a Gaussian kernel of radius = 3 and sigma = 1.5 for better visualization purpose only). The inner 1.5$\arcsec$ radius circle and the outer 8$\arcsec$ circle define the region in between for extracting excess counts in the extended emission. All the images are displayed in logarithmic scale with colors corresponding to number of counts per image pixel. The off-center X-ray source is excluded from our analysis.}
\label{NGC678_chandra}
\end{figure*}

Figure \ref{NGC678_chandra} shows the Chandra images of NGC 678 at the 0.3-3.0 keV and 3.0-7.0 keV bands. The soft X-ray emission appears to be elongated more in the NE-SW direction while the hard band image shows the opposite. An off-center X-ray source is detected $\sim$6.5$\arcsec$ ($\sim$1.3 kpc at the galaxy distance) southwest of the NGC 678 nucleus with an estimated 0.3-7 keV flux of 4.9 $\times$ 10$^{-14}$ erg s$^{-1}$ cm$^{-2}$ and a luminosity of 9.7 $\times$ 10$^{39}$ erg s$^{-1}$ if associated with NGC 678, which could be an ultraluminous X-ray source \citep{Swartz2004}. However, we cannot completely rule out the possibility that this source is part of an ionization cone structure. This off-center X-ray source is excluded from our following analysis of the extended emission. 

\begin{figure*}[h!]
\centering
\includegraphics[width=5.952cm]{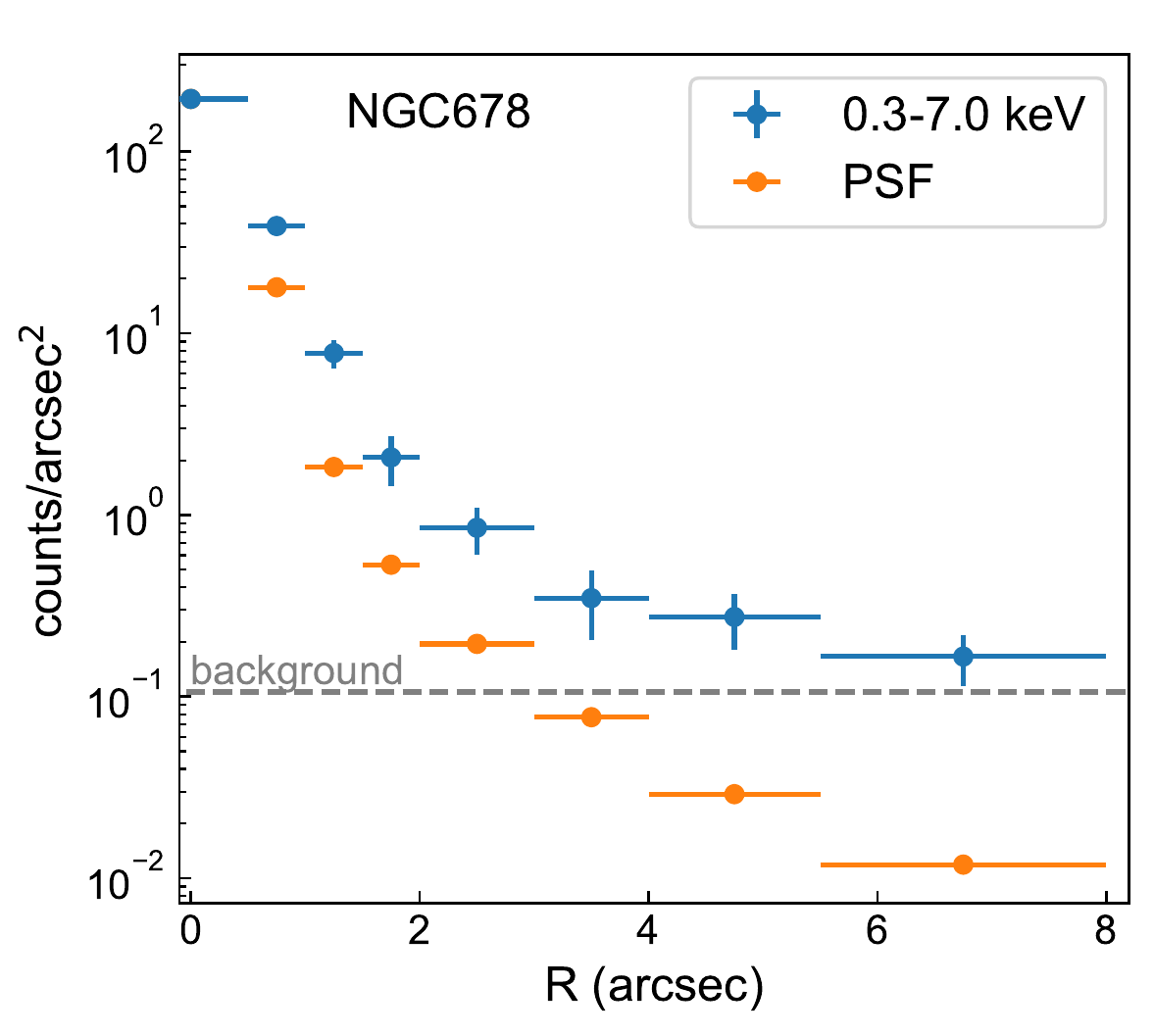}  
\includegraphics[width=5.952cm]{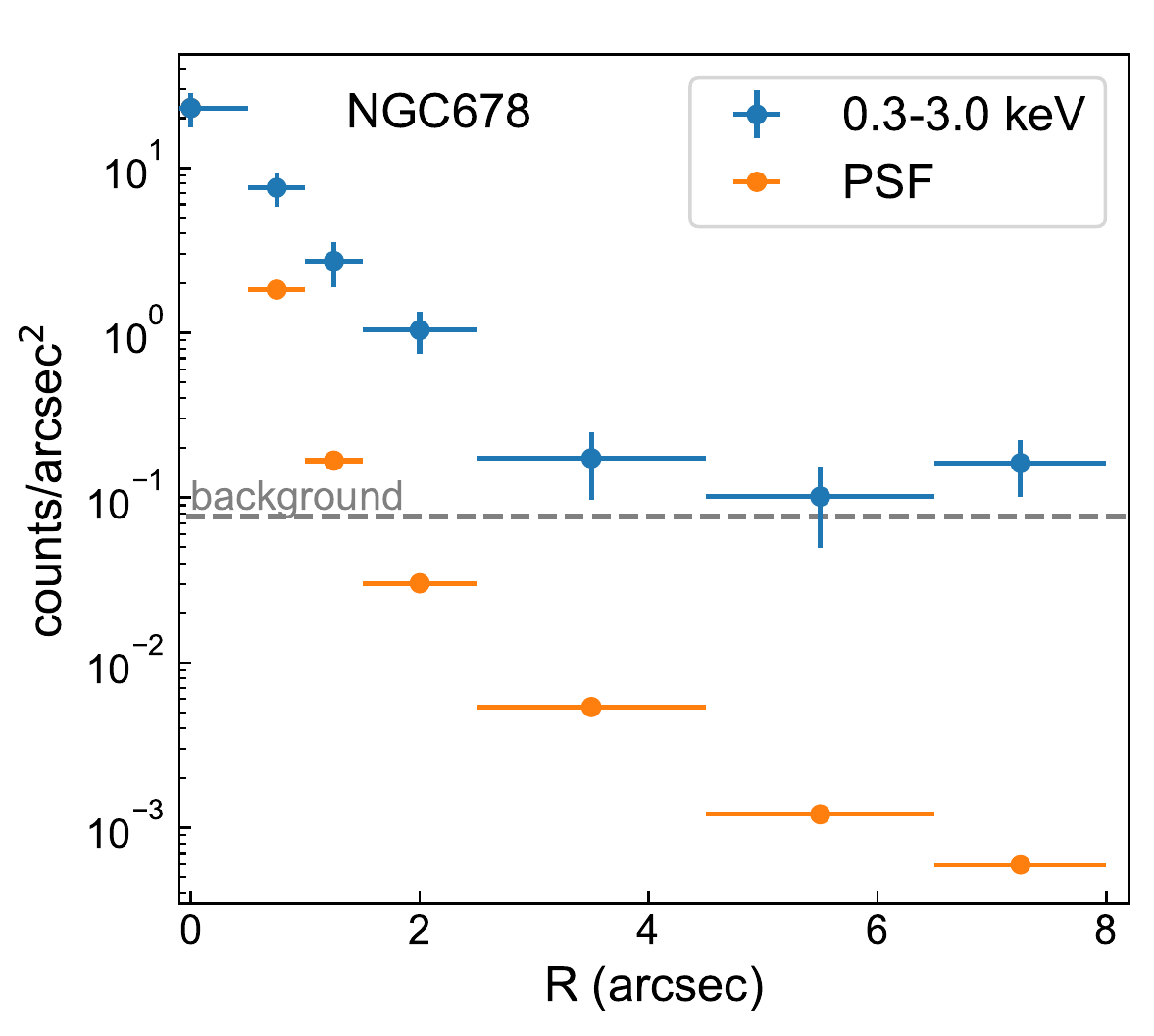}
\includegraphics[width=5.952cm]{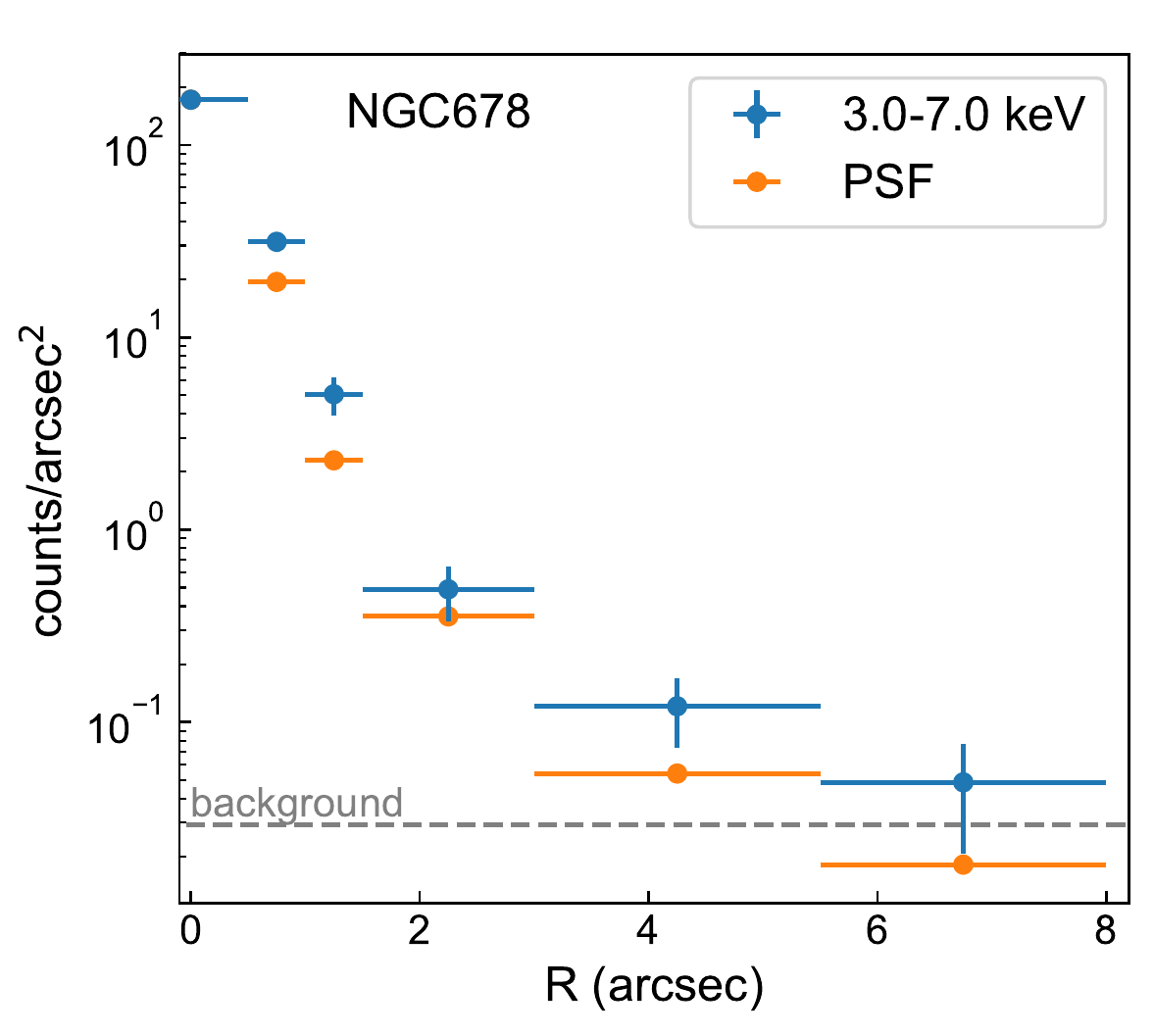}
\caption{Radial profiles of NGC 678 for the full band, soft band, and hard band. The background has been subtracted off from the radial profiles, and the level of which is indicated as the grey dashed horizontal line. The PSF is normalized to the counts in the central 0.5$\arcsec$ radius bin.}
\label{NGC678_radial_profiles}
\end{figure*}

As shown in Figure \ref{NGC678_radial_profiles}, we generate the radial profiles in the full, soft, and hard bands. The soft X-ray emission shows total excess counts well detected at 6.6$\sigma$ above the PSF out to the outer radius (Table \ref{table2}). The 3.0-7.0 keV band radial profile shows excess emission out to at least 5.5$\arcsec$ ($\sim$ 1.1 kpc) with a total of 49.5 $\pm$ 11.1 excess counts detected at 4.4$\sigma$, although the excess counts beyond 1.5$\arcsec$ are not significant (2.0$\sigma$). The total excess fraction in the 3.0-7.0 keV hard band is about 20\%. We also checked the 6.0-7.0 keV band, and the total excess counts in this band are detected above 3$\sigma$ although most of the excess counts are in the inner 0.5$\arcsec$-1.5$\arcsec$ region (out to $\sim$ 300 pc).

\subsection{IC 1657}
\label{IC1657}

IC 1657 is a nearly edge-on ($i$ = 78$^\circ$) barred spiral [SB(s)bc] galaxy at $z$ = 0.01195 (NED; $D$ $\sim$ 53 Mpc; 1$\arcsec$ $\sim$ 252 pc) hosting a Seyfert 2 nucleus \citep{Veron2006} with log$N_{\rm H}$ = 23.40$^{+0.13}_{-0.09}$ cm$^{-2}$ \citep{Ricci2017}. IC 1657 has been studied using optical IFUs \citep{Dopita2015,Lopez2020}. Optical emission line maps reveal a fan-like extended narrow line region emerging mainly from one side (E) perpendicular to the highly inclined N-S galaxy disk. An ionization cone with an opening angle close to 90$^\circ$ is visible as revealed by strong [N\II] emission (Figure 6 in \citealt{Lopez2020}), extending at least $\sim$11$\arcsec$ ($\sim$ 2.8 kpc) from the galaxy disk plane. The galaxy disk itself is dominated by H\II$ $ regions as revealed by the H$\alpha$ emission, while ionized clumps present slightly stronger [O\III] emission at the outskirts of the disk \citep{Lopez2020}.

\begin{figure*} 
\centering
\includegraphics[width=13.5cm]{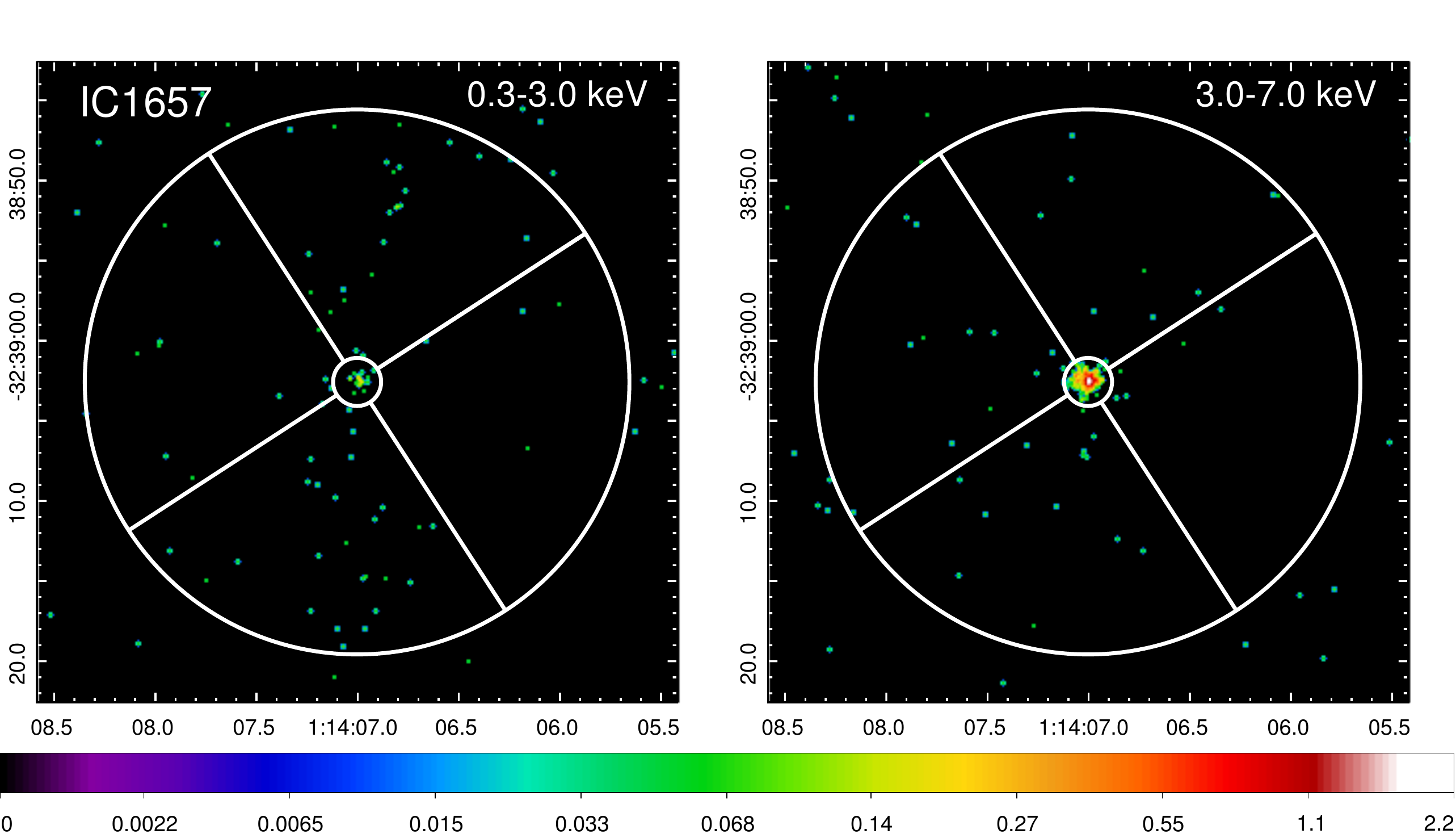}
\caption{40$\arcsec$ $\times$ 40$\arcsec$ Chandra ACIS-S 0.3-3.0 keV (left) and 3.0-7.0 keV (right) band images of IC 1657 at 1/8 subpixel binning (slightly smoothed with a Gaussian kernel of radius = 3 and sigma = 1.5 for better visualization purpose only). The inner 1.5$\arcsec$ radius circle and the outer 17$\arcsec$ circle define the region in between for extracting excess counts in the extended emission. All the images are displayed in logarithmic scale with colors corresponding to number of counts per image pixel.}
\label{IC1657_chandra}
\end{figure*}

\begin{figure*} 
\centering
\includegraphics[width=5.952cm]{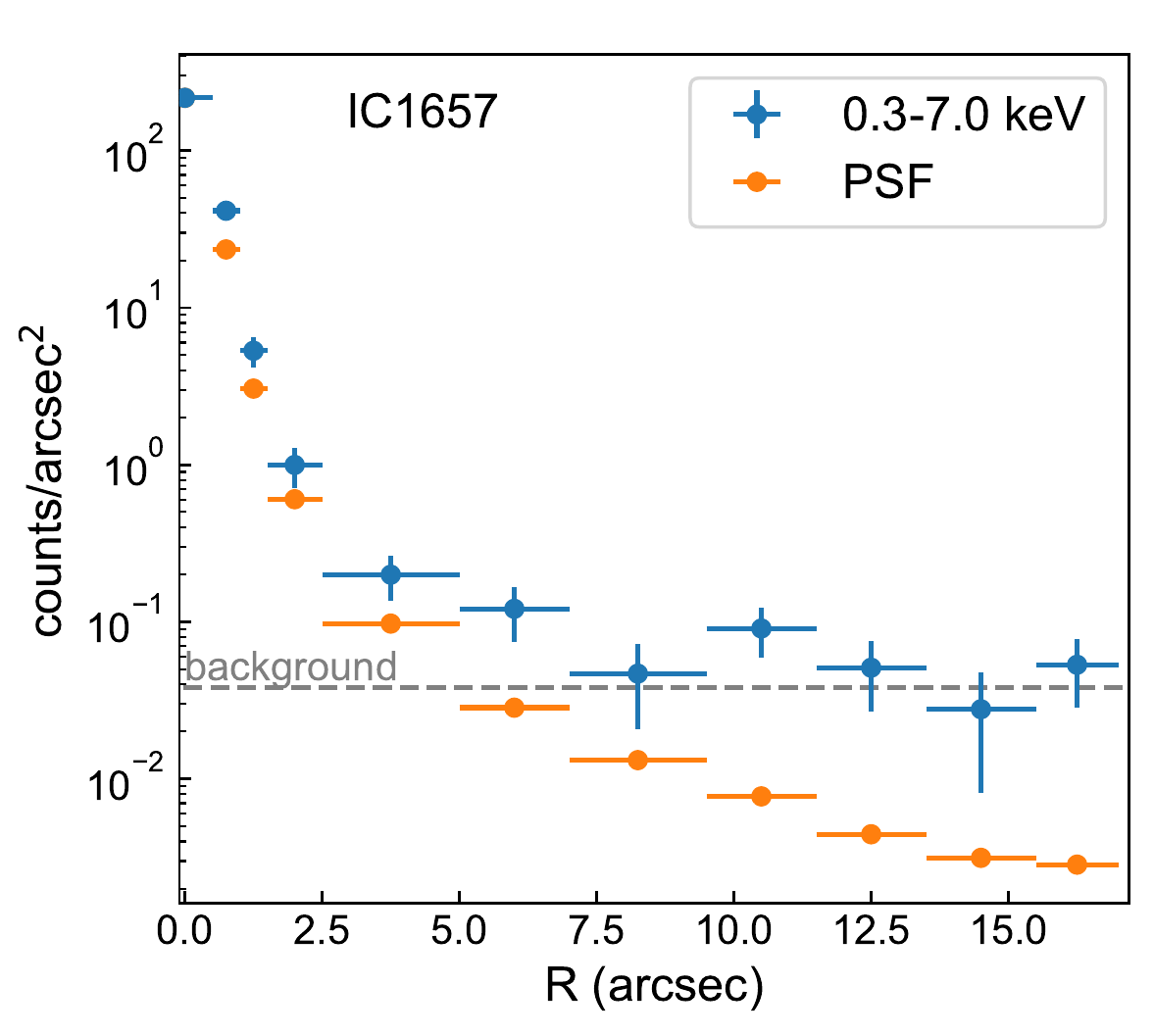}  
\includegraphics[width=5.952cm]{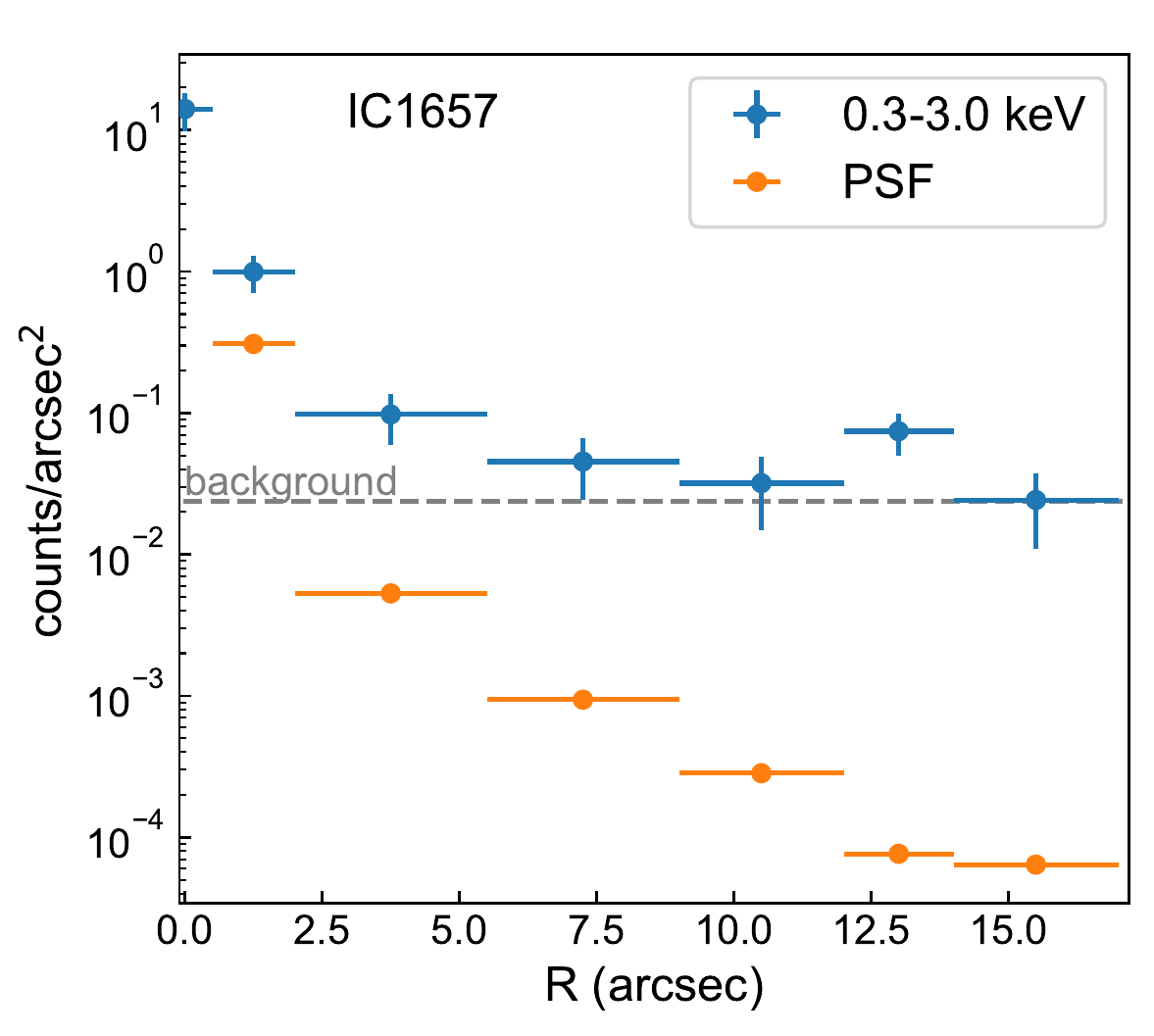}
\includegraphics[width=5.952cm]{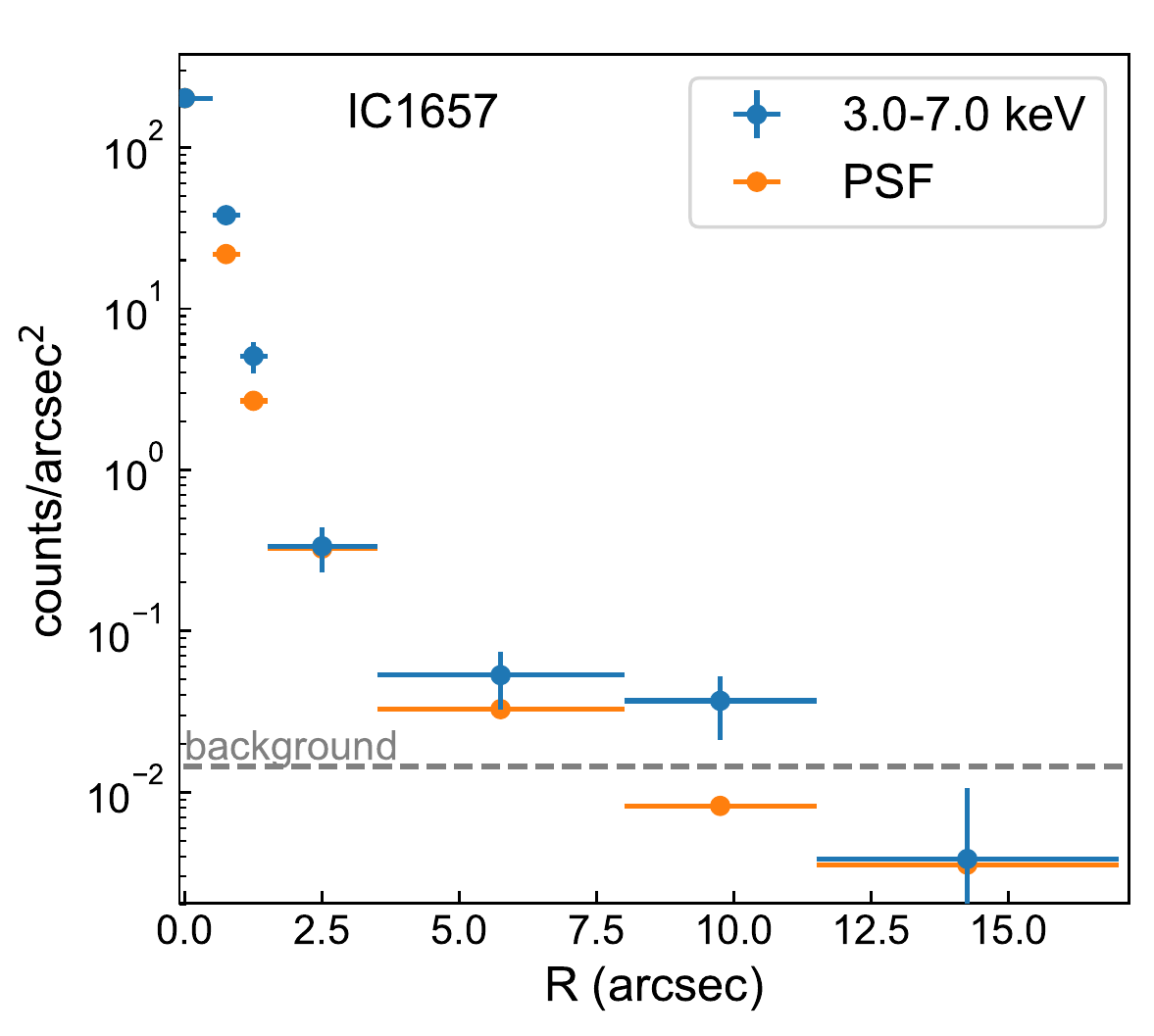}
\caption{Radial profiles of IC 1657 for the full band, soft band, and hard band. The background has been subtracted off from the radial profiles, and the level of which is indicated as the grey dashed horizontal line. The PSF is normalized to the counts in the central 0.5$\arcsec$ radius bin.}
\label{IC1657_radial_profiles}
\end{figure*}

Figure \ref{IC1657_chandra} shows that IC 1657 has a relatively low surface brightness core in the soft band, and the soft X-ray emission is elongated and distributed nearly vertically along the N-S direction, which aligns with the host galaxy disk. There is barely any emission in the perpendicular direction. The excess emission in the soft band is well detected (Table \ref{table2}). The hard band emission does not seem to follow the trend in the soft X-rays. IC 1657 has a total excess counts of 57.7 $\pm$ 12.6 (4.6$\sigma$) in the hard band with about 19\% of the emission in the extended component, but most of the excess emission is found in the inner region out to 1.5$\arcsec$ ($\sim$ 378 pc) (Figure \ref{IC1657_radial_profiles}).

Given the strong azimuthal dependence of the soft X-ray emission, we further divided the region into two azimuthal sectors to examine excess emission in these two sectors separately, the soft X-ray elongation sector and the sector perpendicular to it (cross sector). The extended soft X-ray emission is not detected in the cross sector (Table \ref{table3}). The hard band turns out to have marginally more excess counts in the cross sector with 33.5 $\pm$ 8.8 counts at 3.8$\sigma$ than in the soft X-ray elongation direction with 24.9 $\pm$ 8.7 counts at 2.9$\sigma$, which is contrary to the soft X-ray morphology. 

In most heavily obscured AGN, an ionization cone is present in soft X-rays whenever an optical ionization cone is visible and they normally align with each other, but this is not the case in IC 1657. The optical ionization cone appears perpendicular to the soft X-ray extent, which aligns with the H\II$ $ dominated galaxy disk, and [O\III] is found mostly at the periphery of the star formation region. Instead, as shown earlier, it is the hard X-ray emission that is found perpendicular to, rather than along, the host galaxy plane. We will discuss possible origin of the cone structure and the extended hard X-ray emission in Section \ref{discussion}.

\subsection{NGC 5899}

NGC 5899 is a barred spiral galaxy [SAB(rs)c] \citep{Ann2015} at $z$ = 0.00864 (NED; $\sim$ 38 Mpc; 1$\arcsec$ $\sim$ 183 pc) and hosts a Seyfert 2 nucleus \cite{Veron2006} with log $N_{\rm H}$ = 23.03$^{+0.04}_{-0.03}$ cm$^{-2}$ (\citealt{Ricci2017}; see also \citealt{Balokovic2017}). 

\begin{figure*} 
\centering
\includegraphics[width=13.5cm]{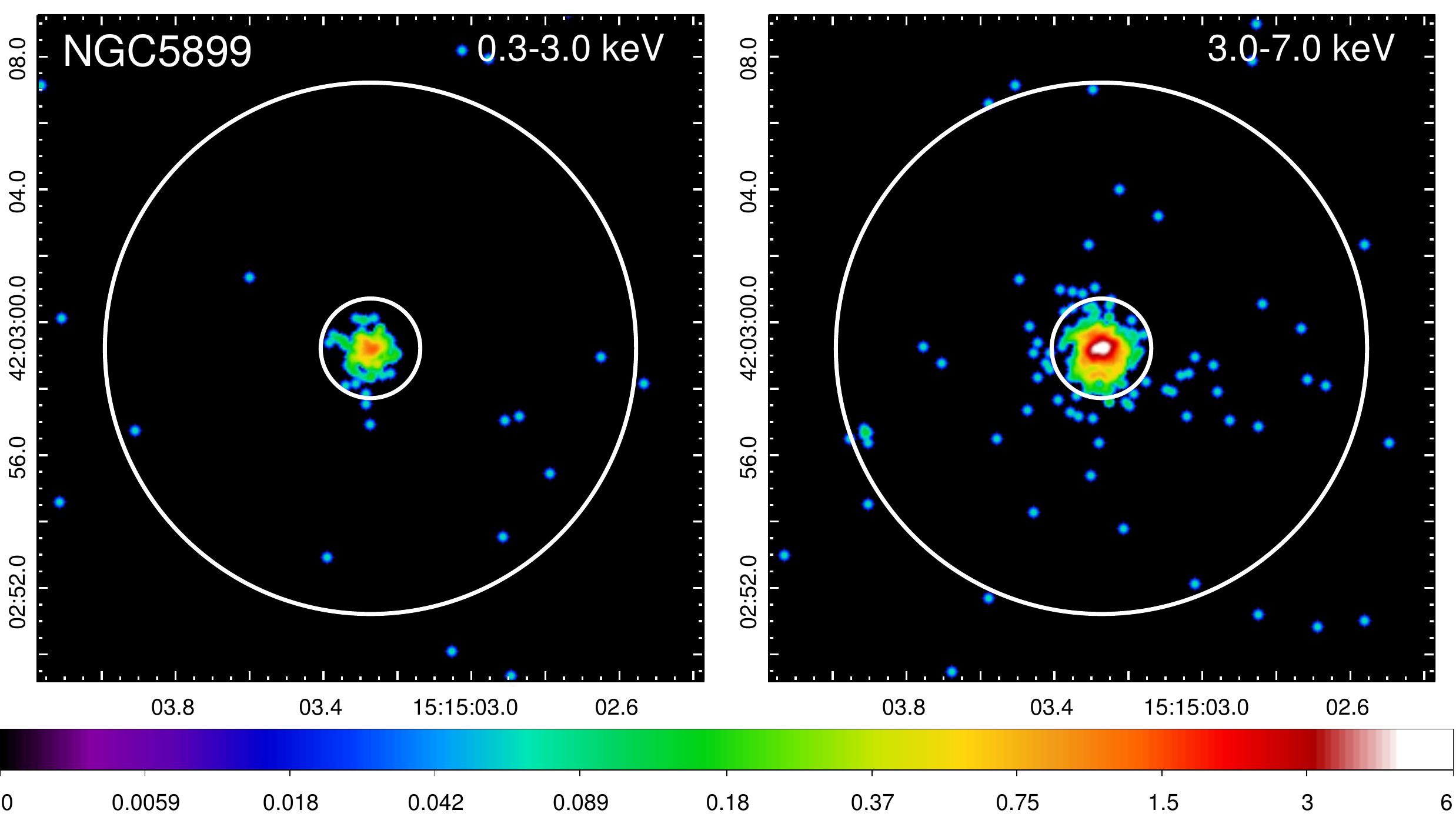}
\caption{20$\arcsec$ $\times$ 20$\arcsec$ Chandra ACIS-S 0.3-3.0 keV (left) and 3.0-7.0 keV (right) band images of NGC 5899 at 1/8 subpixel binning (slightly smoothed with a Gaussian kernel of radius = 3 and sigma = 1.5 for better visualization purpose only). The inner 1.5$\arcsec$ radius circle and the outer 8$\arcsec$ circle define the region in between for extracting excess counts in the extended emission. All the images are displayed in logarithmic scale with colors corresponding to number of counts per image pixel.}
\label{NGC5899_chandra}
\end{figure*}

\begin{figure*} 
\centering
\includegraphics[width=5.952cm]{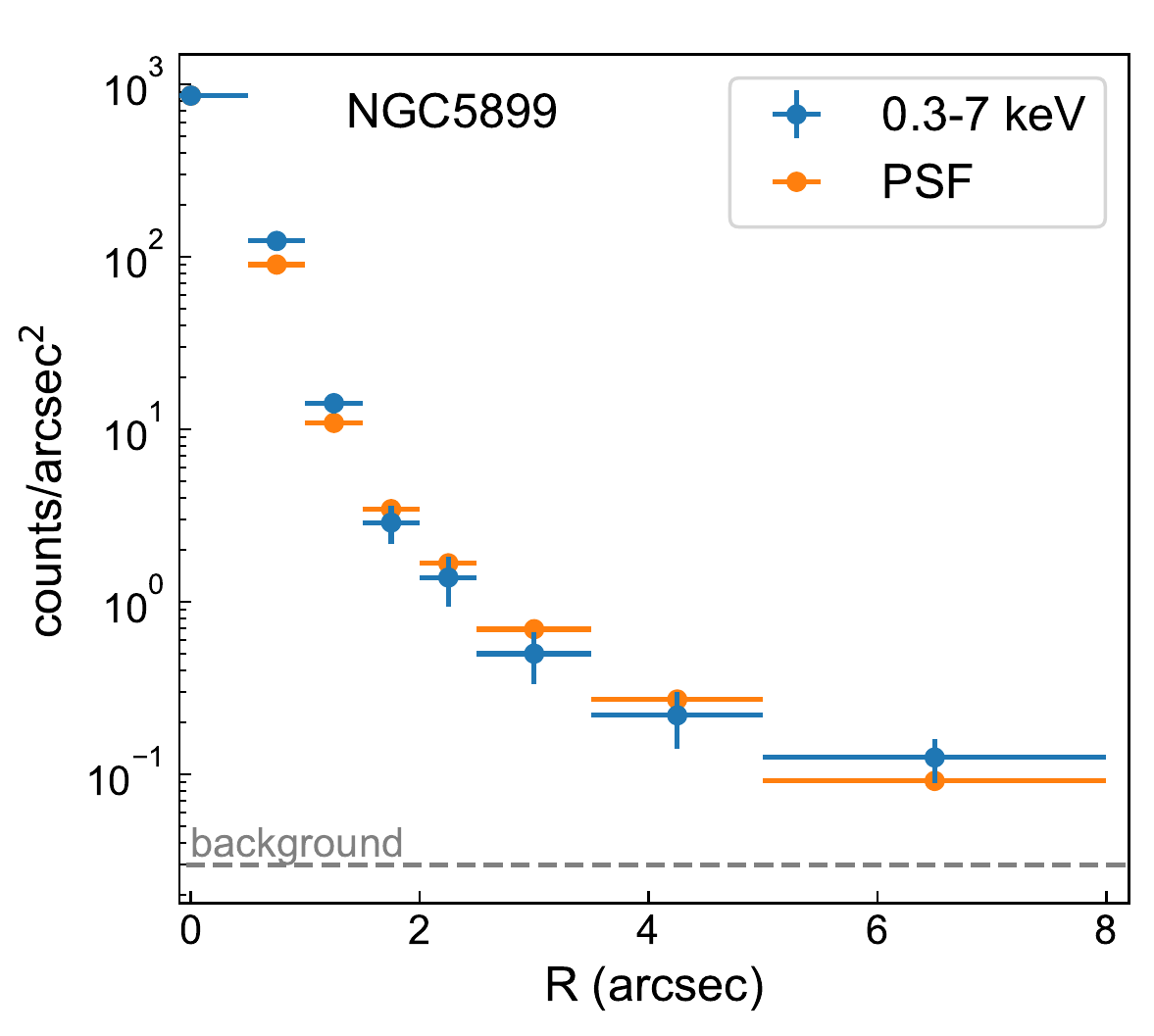}
\includegraphics[width=5.952cm]{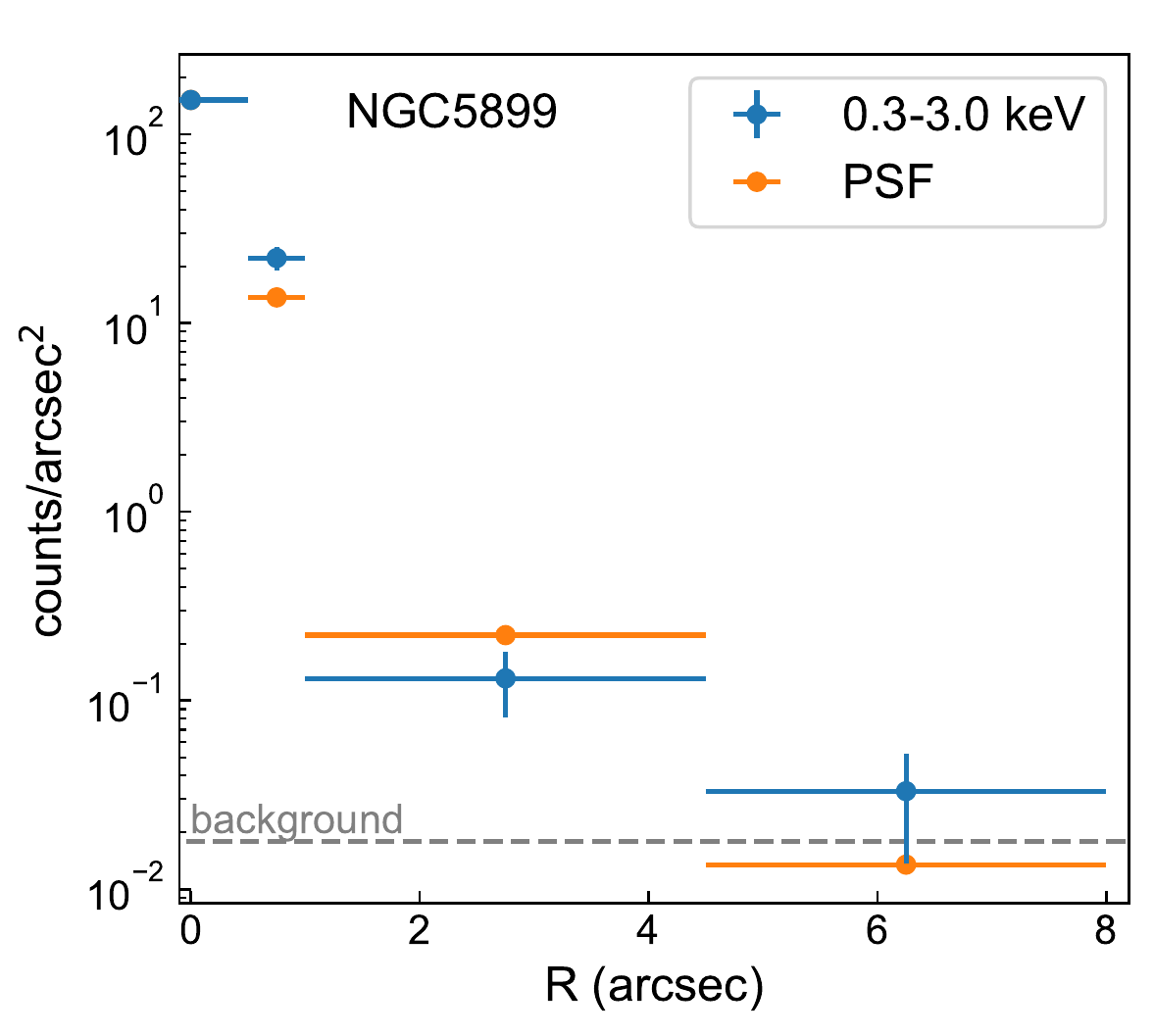}
\includegraphics[width=5.952cm]{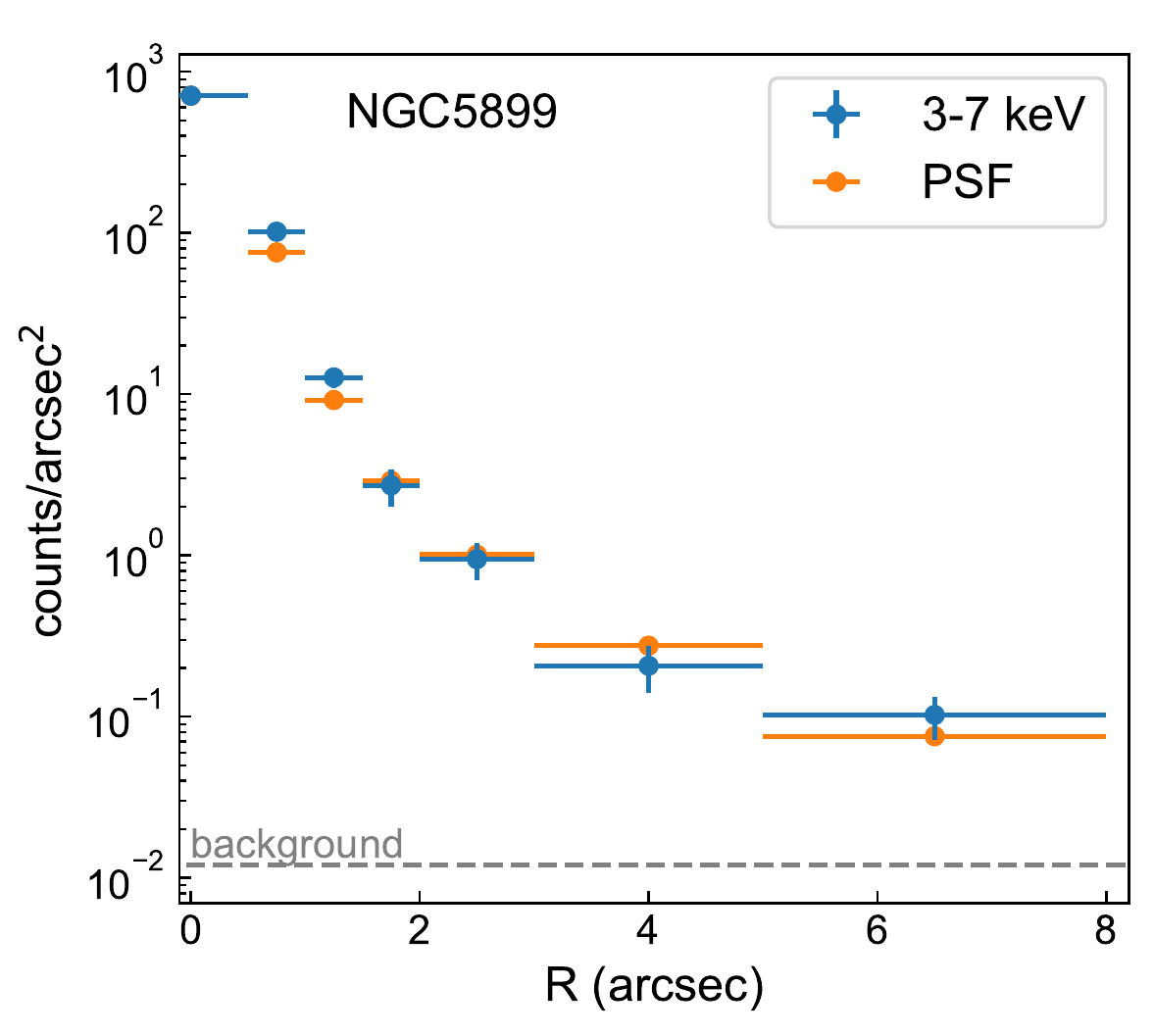}
\caption{Radial profiles of NGC 5899 for the full band, soft band, and hard band. The background has been subtracted off from the radial profiles, and the level of which is indicated as the grey dashed horizontal line. The PSF is normalized to the counts in the central 0.5$\arcsec$ radius bin.}
\label{NGC5899_radial_profiles}
\end{figure*}

As shown in the Chandra images (Figure \ref{NGC5899_chandra}) and radial profiles (Figure \ref{NGC5899_radial_profiles}), NGC 5899 is the only AGN among this sample that does not show extended soft X-ray emission. In the 3.0-7.0 keV band, there are some excess counts detected at 3.9$\sigma$ above the PSF with a total excess fraction of $\sim$ 8\% (Table \ref{table2}), but beyond 1.5$\arcsec$ ($\sim$ 275 pc) the radial profile is basically consistent with the PSF (i.e., no excess emission). 

\subsection{NGC 454E}

NGC 454E is an early-type galaxy at $z$ = 0.01213 (NED; $D$ $\sim$ 54 Mpc; 1$\arcsec$ $\sim$ 255 pc) in a pair of interacting galaxies \citep{Johansson1988,Stiavelli1998} with log$N_{\rm H}$ = 23.30$^{+0.04}_{-0.03}$ cm$^{-2}$ \citep{Ricci2017}. NGC 454E is also identified as a new member of the class of Ôchanging-look AGN' ($N_{\rm H}$ varying from $\sim$ 1 $\times$ 10$^{24}$ cm$^{-2}$ to $\sim$ 1 $\times$ 10$^{23}$ cm$^{-2}$; \citealt{Marchese2012,Balokovic2017}), i.e., AGN that show significant variation of the absorbing column density along the line of sight \citep{Matt2003}.

\begin{figure*} 
\centering
\includegraphics[width=13.5cm]{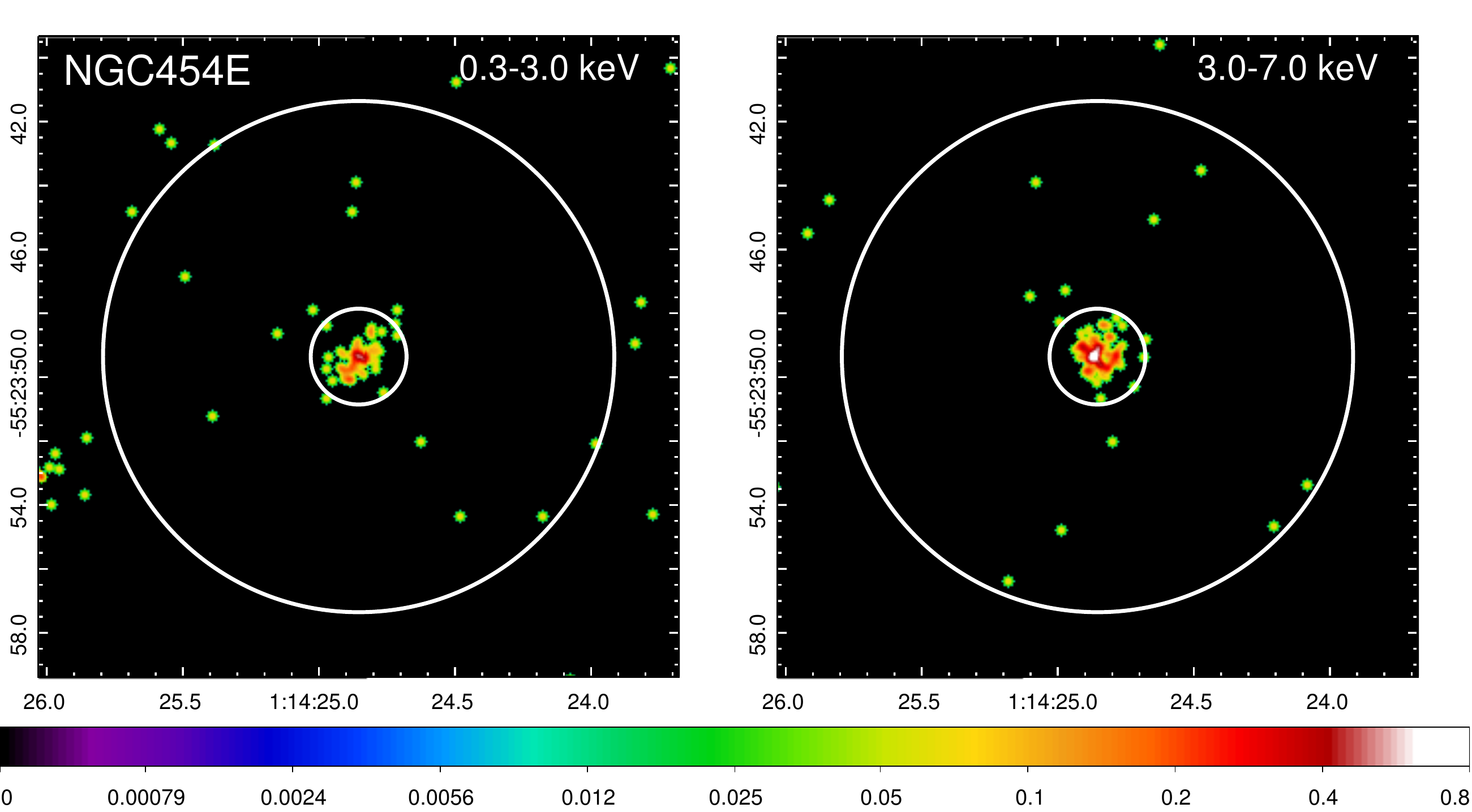}
\caption{20$\arcsec$ $\times$ 20$\arcsec$ Chandra ACIS-S 0.3-3.0 keV (left) and 3.0-7.0 keV (right) band images of NGC 454E at 1/8 subpixel binning (slightly smoothed with a Gaussian kernel of radius = 3 and sigma = 1.5 for better visualization purpose only). The inner 1.5$\arcsec$ radius circle and the outer 8$\arcsec$ circle define the region in between for extracting excess counts in the extended emission. All the images are displayed in logarithmic scale with colors corresponding to number of counts per image pixel.}
\label{NGC454E_chandra}
\end{figure*}

\begin{figure*} 
\centering
\includegraphics[width=5.952cm]{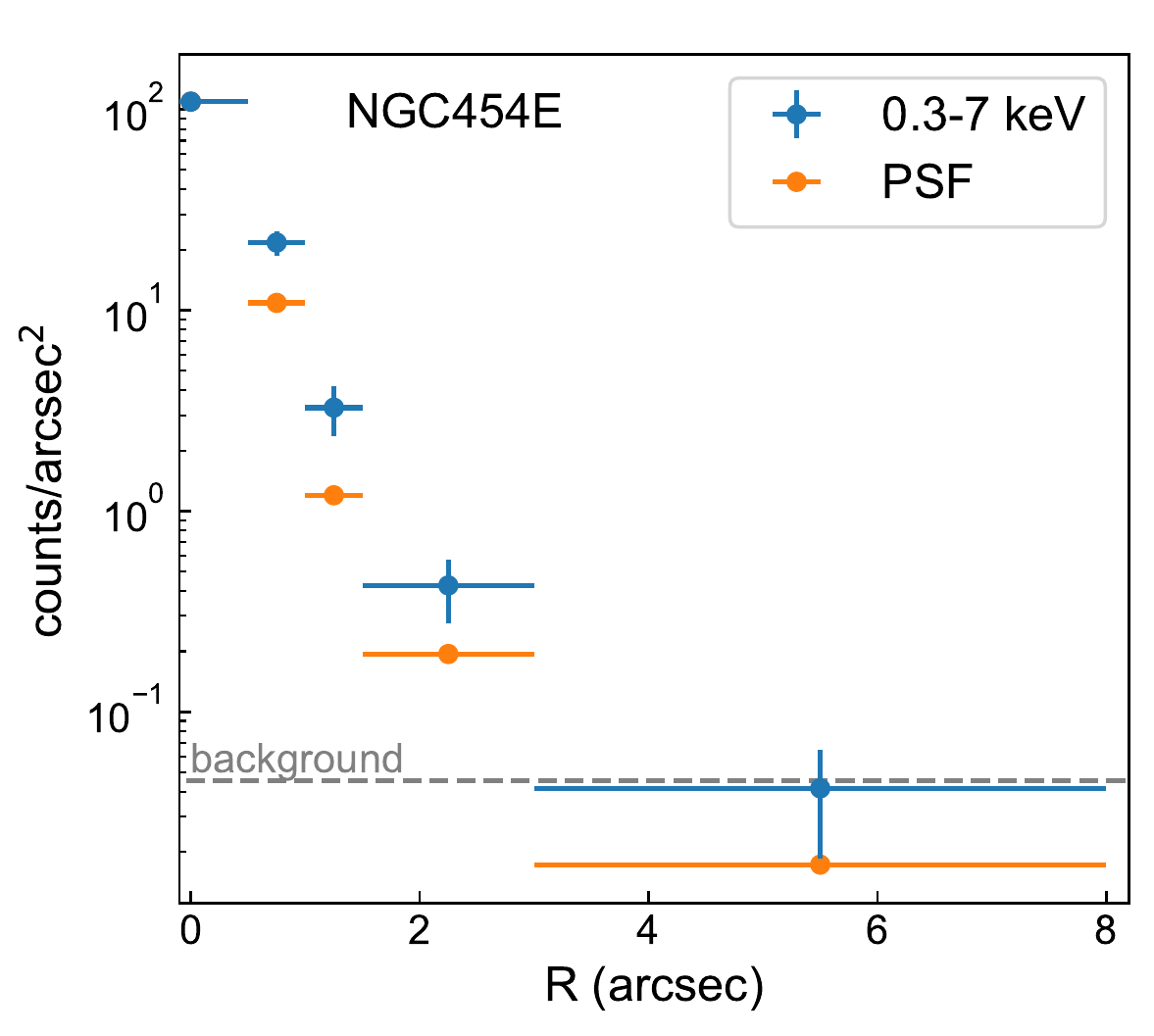}
\includegraphics[width=5.952cm]{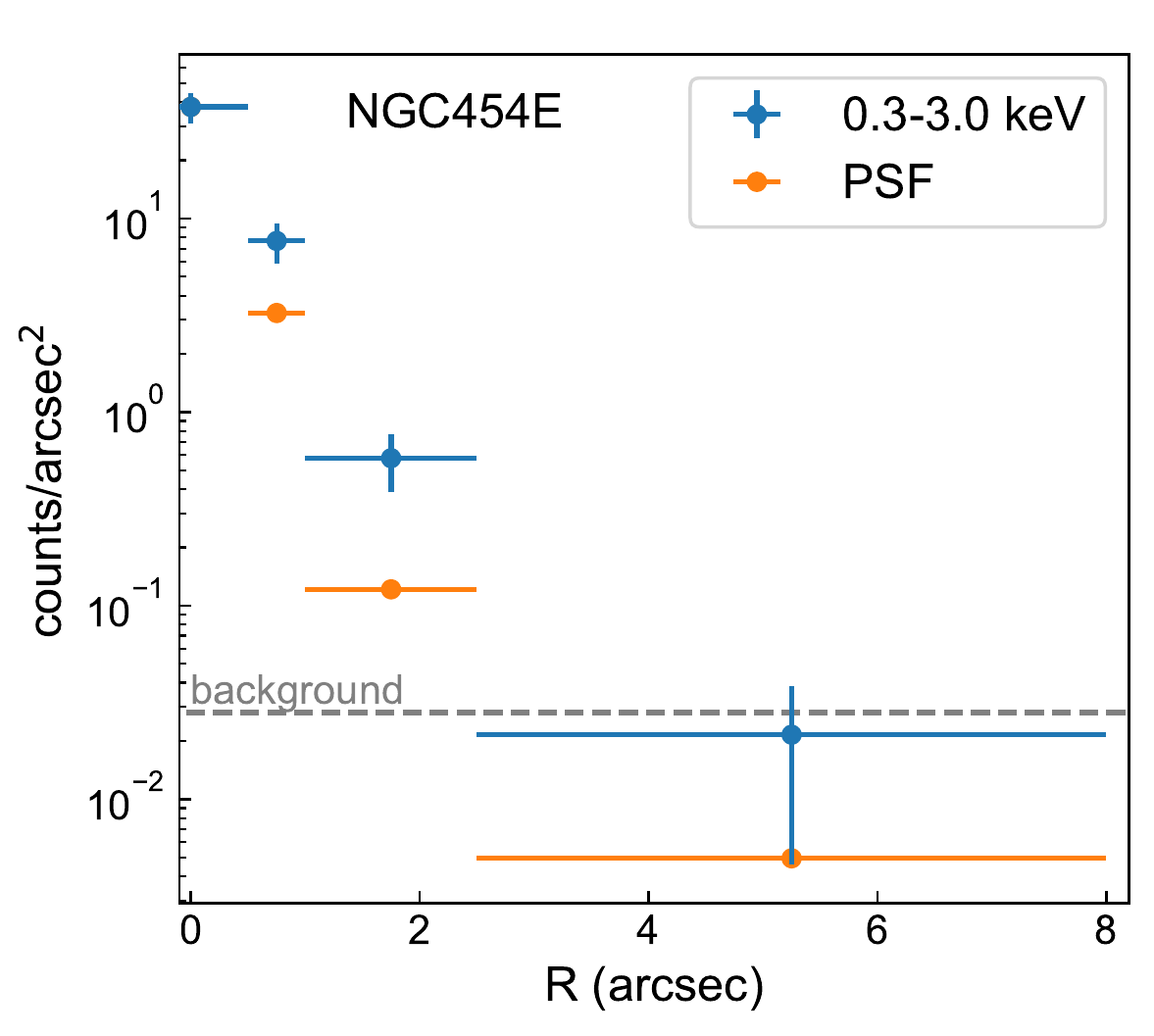}
\includegraphics[width=5.952cm]{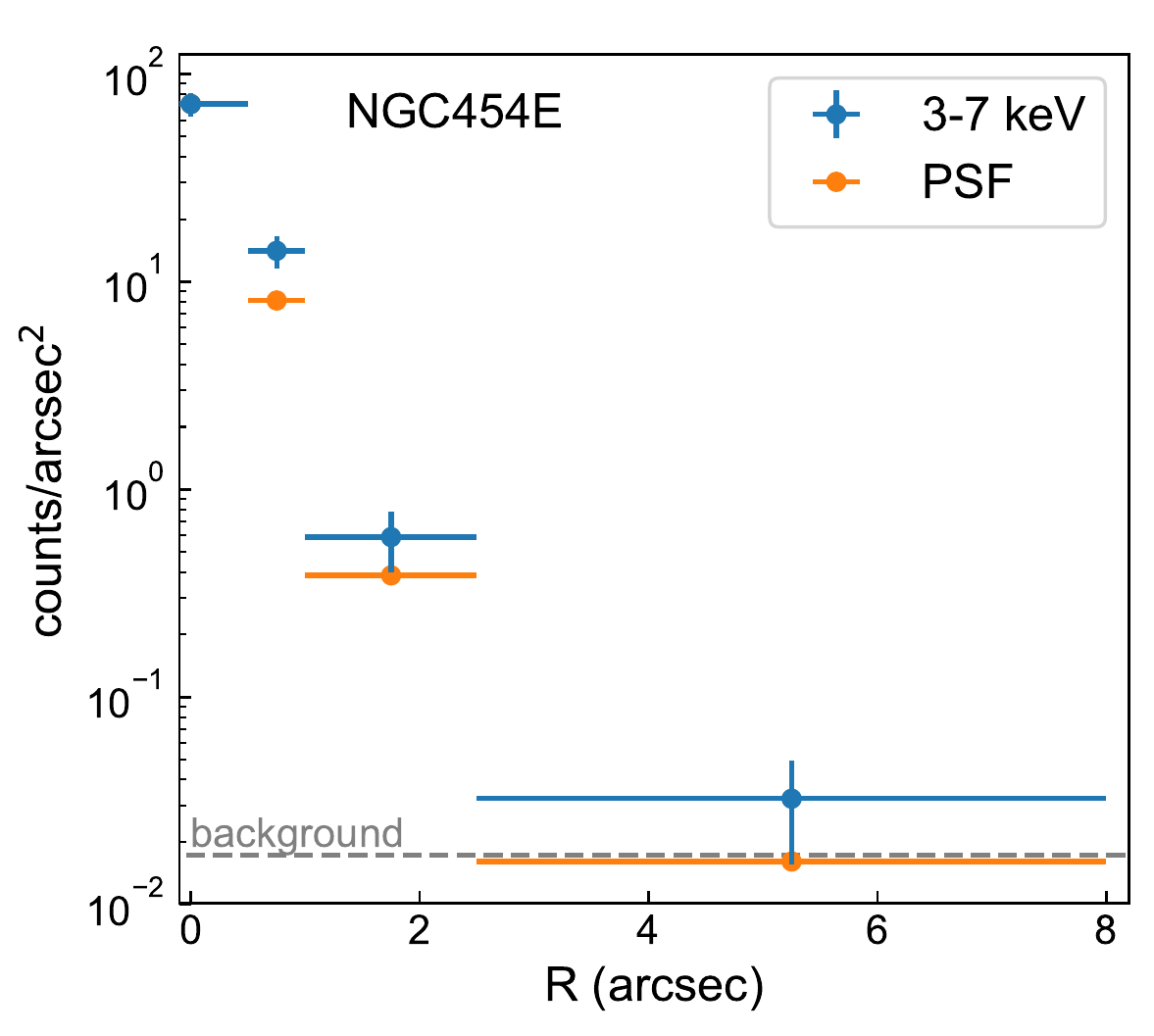}
\caption{Radial profiles of NGC 454E for the full band, soft band, and hard band. The background has been subtracted off from the radial profiles, and the level of which is indicated as the grey dashed horizontal line. The PSF is normalized to the counts in the central 0.5$\arcsec$ radius bin.}
\label{NGC454E_radial_profiles}
\end{figure*}

The soft X-ray emission in NGC 454E shows some excess counts above the PSF mostly in the inner 0.5-1.0$\arcsec$ region (Figures \ref{NGC454E_chandra} and \ref{NGC454E_radial_profiles}). In the hard band, there are also some excess counts in the inner 0.5-1.0$\arcsec$ region but they are not statistically significant, and extended emission beyond 1.5$\arcsec$ is not detected (Table \ref{table2}). 

\subsection{ESO 234-G050}
\label{ESO234-G050}

ESO 234-G050 appears to be a low surface brightness spiral galaxy\footnote{ESO 234-G050 was classified as a blue compact dwarf (BCD) elliptical galaxy by \cite{Aguero1993}.} ($z$ = 0.00877; $D$ $\sim$ 39 Mpc; 1$\arcsec$ $\sim$ 185 pc; NED) in the DECam Legacy Survey imaging \citep{Dey2019}. It hosts a Seyfert 2 nucleus \citep{Aguero1993,Onori2017} with log $N_{\rm H}$ = 23.08$^{+0.12}_{-0.18}$ cm$^{-2}$ \citep{Ricci2017}.

\begin{figure*} 
\centering
\includegraphics[width=13.5cm]{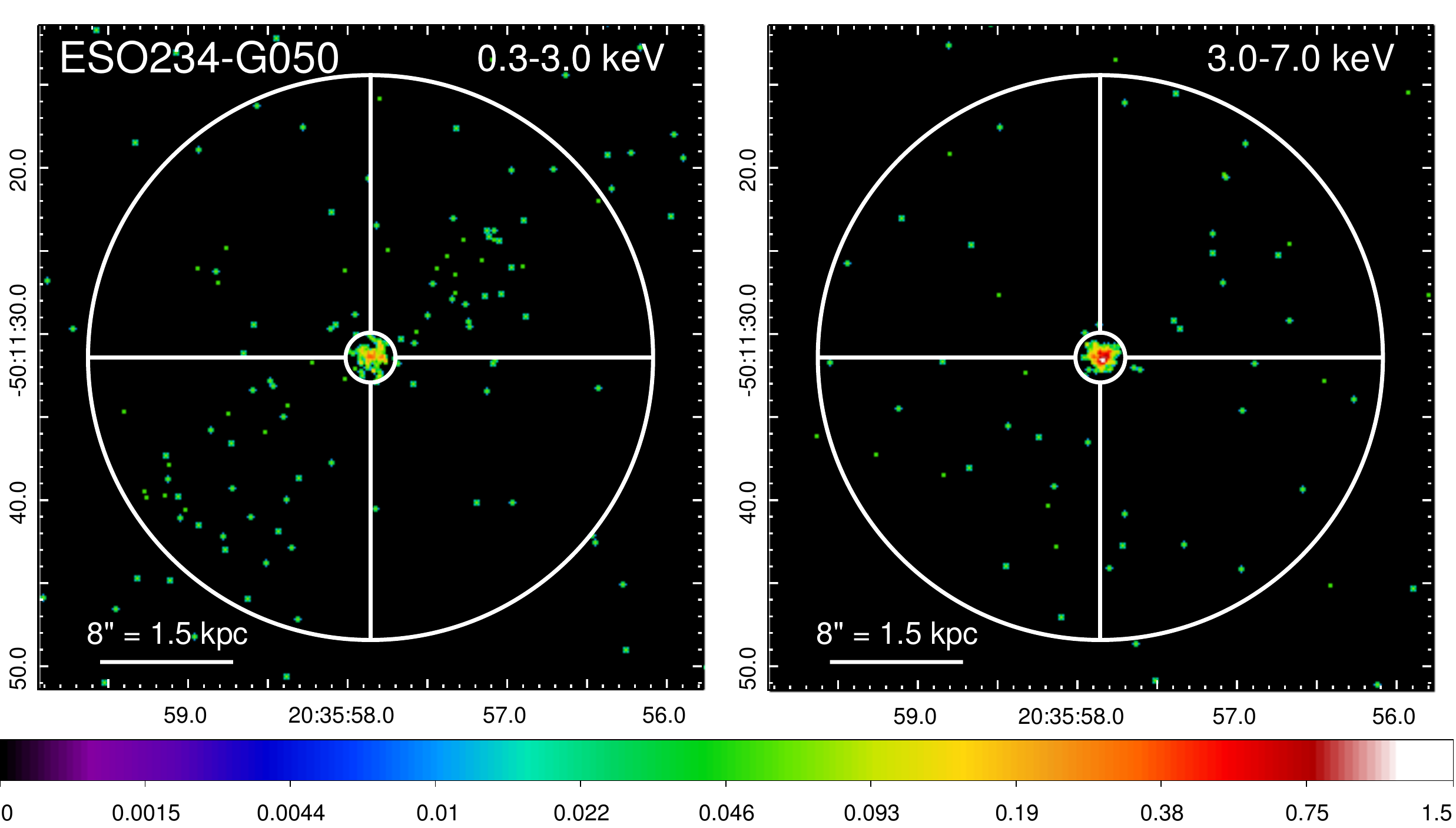}
\caption{40$\arcsec$ $\times$ 40$\arcsec$ Chandra ACIS-S 0.3-3.0 keV (left) and 3.0-7.0 keV (right) band images of ESO 234-G050 at 1/8 subpixel binning (slightly smoothed with a Gaussian kernel of radius = 3 and sigma = 1.5 for better visualization purpose only). The inner 1.5$\arcsec$ radius circle and the outer 15$\arcsec$ circle define the region in between for extracting excess counts in the extended emission. All the images are displayed in logarithmic scale with colors corresponding to number of counts per image pixel.}
\label{ESO234-G050_chandra}
\end{figure*}

\begin{figure*} 
\centering
\includegraphics[width=5.952cm]{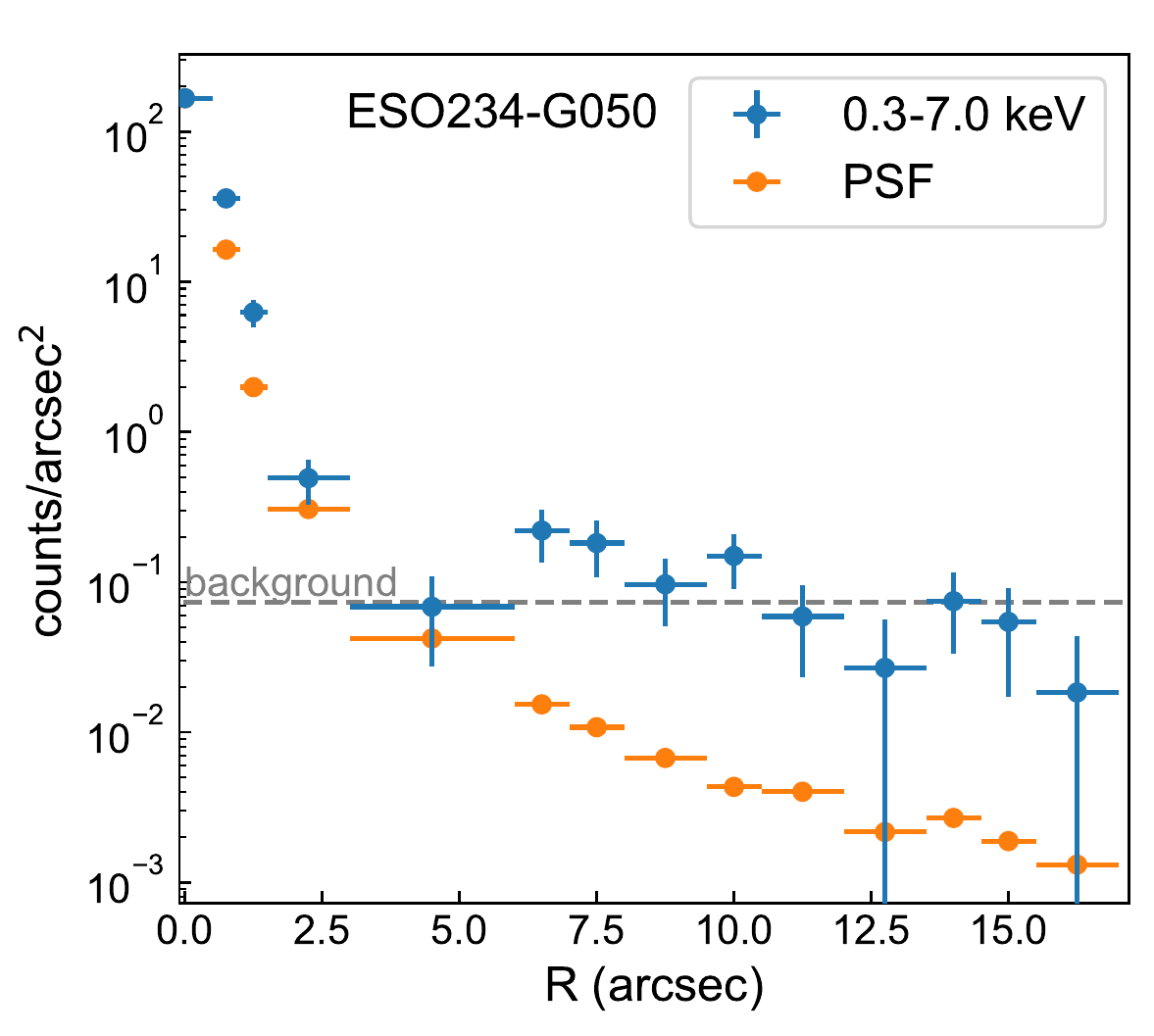} 
\includegraphics[width=5.952cm]{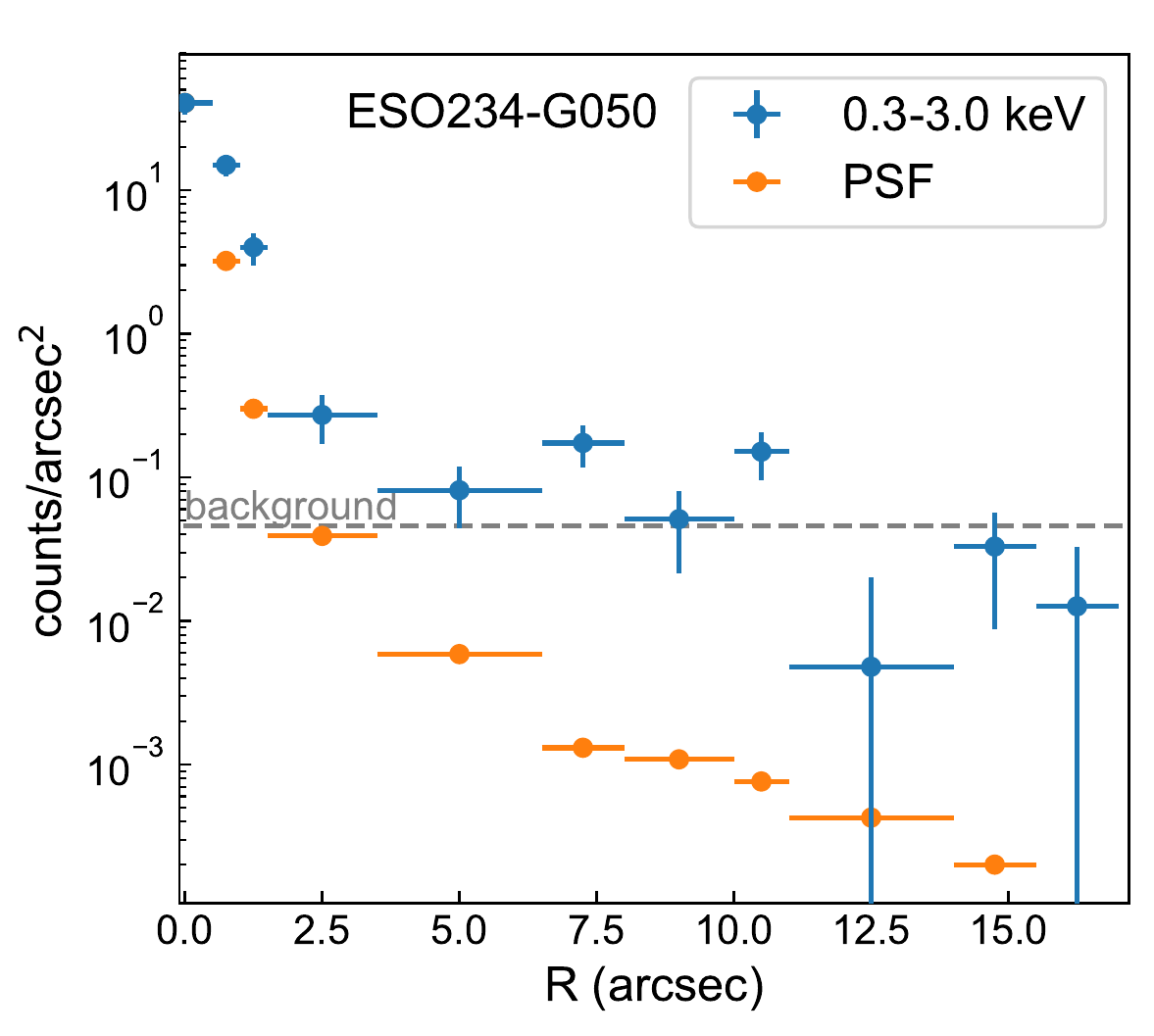}
\includegraphics[width=5.952cm]{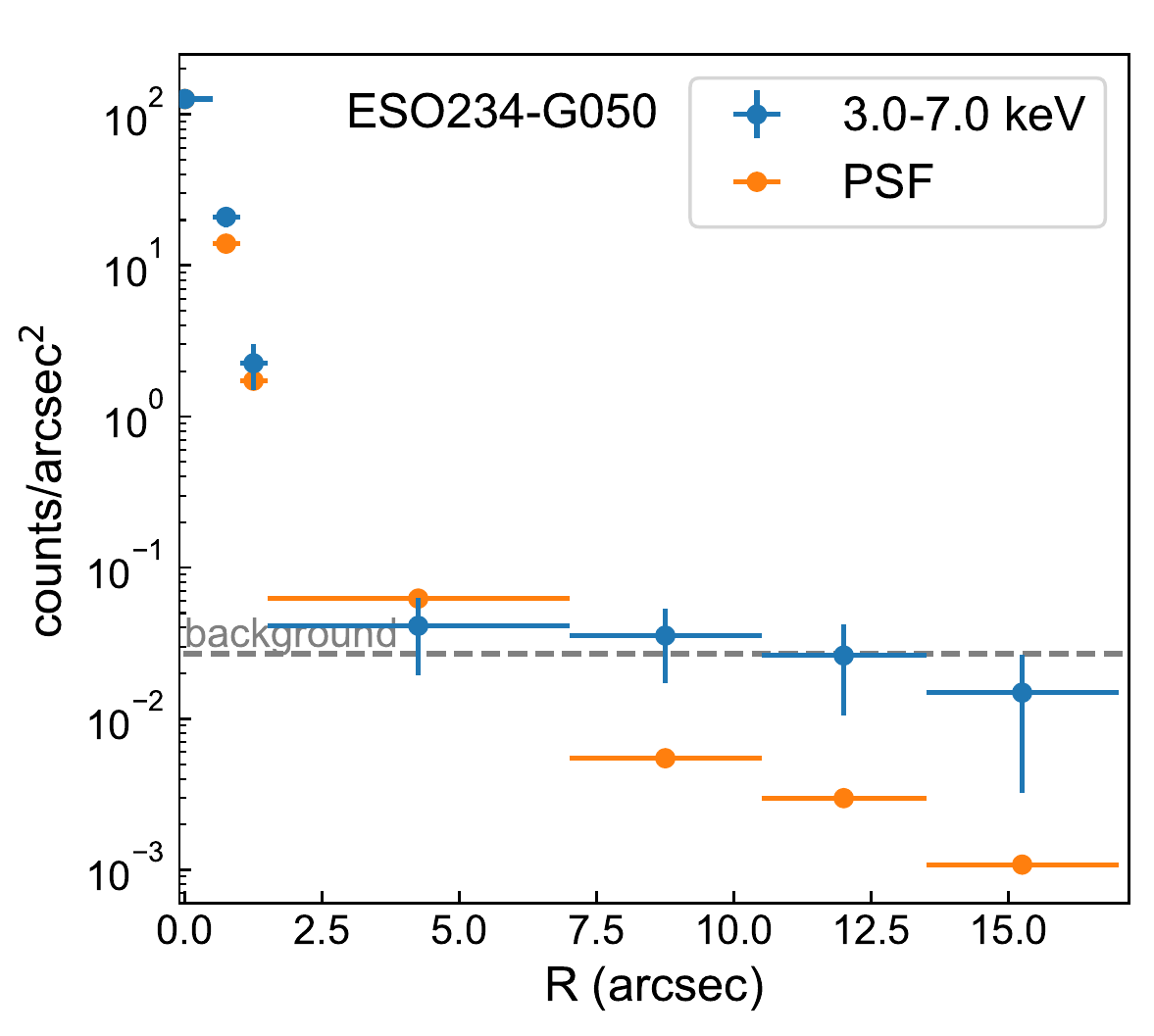}
\caption{Radial profiles of ESO 234-G050 for the full band, soft band, and hard band. The background has been subtracted off from the radial profiles, and the level of which is indicated as the grey dashed horizontal line. The PSF is normalized to the counts in the central 0.5$\arcsec$ radius bin.}
\label{ESO234-G050_radial_profiles}
\end{figure*}

ESO 234-G050 exhibits prominent extended soft X-ray emission preferentially distributed along the NW-SE direction as shown in Figure \ref{ESO234-G050_chandra}, which resembles an ionization bicone structure. The hard X-ray band has a concentrated surface brightness distribution in the center (Figures \ref{ESO234-G050_chandra} and \ref{ESO234-G050_radial_profiles}), but no extended emission is detected above 3$\sigma$. Since the soft X-ray emission has a strong azimuthal dependence, we split the data into two azimuthal sectors, one in the NW-SE direction along the soft X-ray elongation and one in the NE-SW direction, to examine excess emission in each sector separately. ESO 234-G050 does not have enough counts for us to produce radial profiles for each azimuthal sector though. The extended soft X-ray emission in the soft X-ray elongation sector is well detected as expected. There is also some excess emission detected in the cross sector, mostly in the inner 0.5$\arcsec$-1.5$\arcsec$ region. For the hard band, there are slightly more excess counts in the soft X-ray elongation sector than in the cross sector. However, none of them is detected above 3$\sigma$.

\section{Discussion}
\label{discussion}

\begin{table*}
\centering
\caption{Estimated fluxes and luminosities of the measured extended components. }
\begin{tabular}{lcccccc}
\hline\hline
Sourcename &  $N_{\rm H}$ & $f_{\rm 0.3-3 keV}$  &    $L_{\rm 0.3-3 keV}$       &        $f_{\rm 3-7 keV}$&    $L_{\rm 3-7 keV}$           \\
                      &    10$^{20}$ cm$^{-2}$&10$^{-14}$ erg s$^{-1}$ cm$^{-2}$ & 10$^{39}$ erg s$^{-1}$            &10$^{-14}$ erg s$^{-1}$ cm$^{-2}$ & 10$^{39}$ erg s$^{-1}$ \\
\hline
NGC 678			&7.57 	& 1.8  &3.5   &2.9  &5.7  \\
IC 1657           		&2.30	& 4.0  &12.9  & 8.7 & 27.8  \\
NGC 5899		&1.77	& 2.2  & 3.6  &16.4  &27.1\\
NGC 454E		&2.72	& 1.7  & 5.7  & 3.1 &10.1\\
ESO 234-G050  	&3.33	& 5.6  & 9.6  & 3.5 & 6.0 \\
\hline
NGC 424       		&1.76	& 18.9  &58.2   &5.8  & 17.9  \\
NGC 1125			&2.73	& 4.7  &12.6  & 1.6 & 4.1 \\
NGC 3281		&6.49	& 7.6  & 19.2   & 11.4  & 28.9 \\
NGC 4500		&1.05	& 6.1  &14.5   & 2.6 & 6.2 \\
ESO 005-G004		&11.54	& 0.7 & 0.6 & 2.9 &2.5 \\
ESO 137-G034		&24.95	& 11.1 & 20.0  & 4.6 & 8.3 \\
2MASXJ00253292+6821442 & 51.31	& 0.4  &1.4   & 1.1&3.5 \\
\hline
\hline
\end{tabular}
\label{table4}
\tablecomments{$N_{\rm H}$ is the Galactic absorption column density. We assume an absorbed power-law model with a photon index of 1.9.}
\end{table*}

\begin{table*}
\centering
\caption{Expected X-ray luminosities from XRBs}
\begin{tabular}{lccccccccc}
\hline\hline
Sourcename & Morphological  & $L_{\rm 0.3-3 keV}^{\rm LMXB}$ &$L_{\rm 3-7 keV}^{\rm LMXB}$& $L_{\rm 0.3-3 keV}^{\rm HMXB/mixed}$ &$L_{\rm 3-7 keV}^{\rm HMXB/mixed}$& $r_{\rm 0.3-3 keV}$ & $r_{\rm 3-7 keV}$   & $f_{\rm 3-7 keV}^{\rm cross-cone}$  \\
                      &    type   &10$^{39}$ erg s$^{-1}$            & 10$^{39}$ erg s$^{-1}$ &10$^{39}$ erg s$^{-1}$            & 10$^{39}$ erg s$^{-1}$ &\\
\hline
NGC 678			&SB(s)b; edge-on	&2.3 &1.9 &          &      & 65\% & 33\%\\
IC 1657           		&SB(s)bc	&3.2	 & 2.4 &7.5 &5.4 &25\%-58\% &9\%-19\%  &13\%-29\%   \\
NGC 5899		&SAB(rs)c	&1.4 &1.0 &          &      & 39\% & 4\%\\
NGC 454E		&E	        &2.7 &2.0&          &      &47\% & 20\%\\
ESO 234-G050  	&S &0.4& 0.5&          &      & 4\% & 8\% &13\%\\
\hline
NGC 424       		&SB0/a	&1.3& 1.0   & 5.0 & 3.6 & 2\%-9\%  & 6\%-20\%\\
NGC 1125			&SB0/a	&0.5 & 0.4  &2.1 & 1.5  & 4\%-17\% & 12\%-36\% \\
NGC 3281		&SAab	&2.3& 1.7  & 7.6& 5.5  & 12\%-40\% & 6-19\% & 9\%-29\%\\
NGC 4500		&SBa	&1.6& 1.2 &          &      &11\% & 20\%\\
ESO 005-G004		&Sb; edge-on   &1.2 &0.9&          &      &214\% & 36\% \\
ESO 137-G034		&S0/a	&5.1 & 3.9 &3.5  &2.5 &18\%-26\% & 30\%-47\% & 45\%-70\% \\
J0025+6821 & ---&1.5&1.2&          &      &110\% & 34\% \\
\hline
\hline
\end{tabular}
\label{table5}
\tablecomments{J0025+6821 is a short name for 2MASXJ00253292+6821442. $L_{\rm 0.3-3 keV}^{\rm LMXB}$ and $L_{\rm 3-7 keV}^{\rm LMXB}$ are derived from the scaling relation of LMXBs in \cite{Boroson2011}. For NGC 424, NGC 1125, and NGC 3281, $L_{\rm 0.3-3 keV}^{\rm HMXB}$ and $L_{\rm 3-7 keV}^{\rm HMXB}$ are derived from the scaling relation of HMXBs in \cite{Mineo2012,Mineo2014}. For IC 1657 and ESO 137-G034, $L_{\rm 0.3-3 keV}^{\rm mixed}$ and $L_{\rm 3-7 keV}^{\rm mixed}$ are derived based on the relation with both SFR and stellar mass from \cite{Lehmer2019}. $r_{\rm 0.3-3 keV}$ and $r_{\rm 3-7 keV}$ are the percentage ratios of the XRB expected to the measured $L_{\rm 0.3-3 keV}$ and $L_{\rm 3-7 keV}$, respectively. $f_{\rm 3-7 keV}^{\rm cross-cone}$ is the fraction of the cross-cone excess emission that could be explained by the expected XRBs.  }
\end{table*}

\begin{figure} 
\centering
\includegraphics[width=9cm]{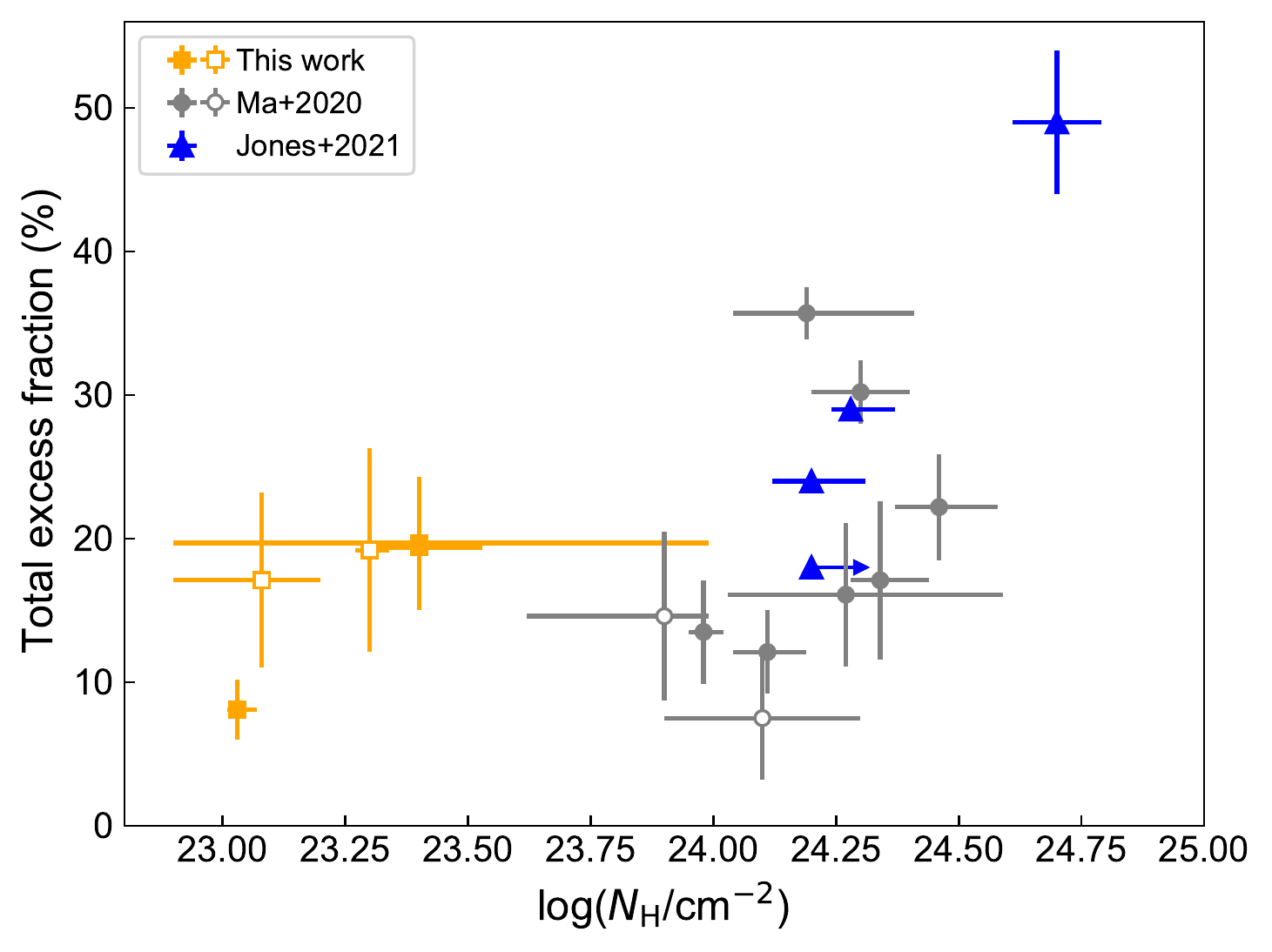}  
\caption{3.0-7.0 keV total excess fraction versus log$N_{\rm H}$. The orange squares are the five obscured AGN in this work. The gray circles are the seven CT AGN plus two individually selected CT AGN (ESO 428-G014 and NGC 7212) in \cite{Ma2020}. For the open squares or circles, the total excess fractions are below 3$\sigma$. The blue triangles are the CT AGN from \cite{Jones2021} with the excess fractions measured in the ionization cone direction.}
\label{Nh_excfrac}
\end{figure}

\subsection{Potential contributions from X-ray binaries}

\begin{figure*} 
\centering
\includegraphics[width=8.5cm]{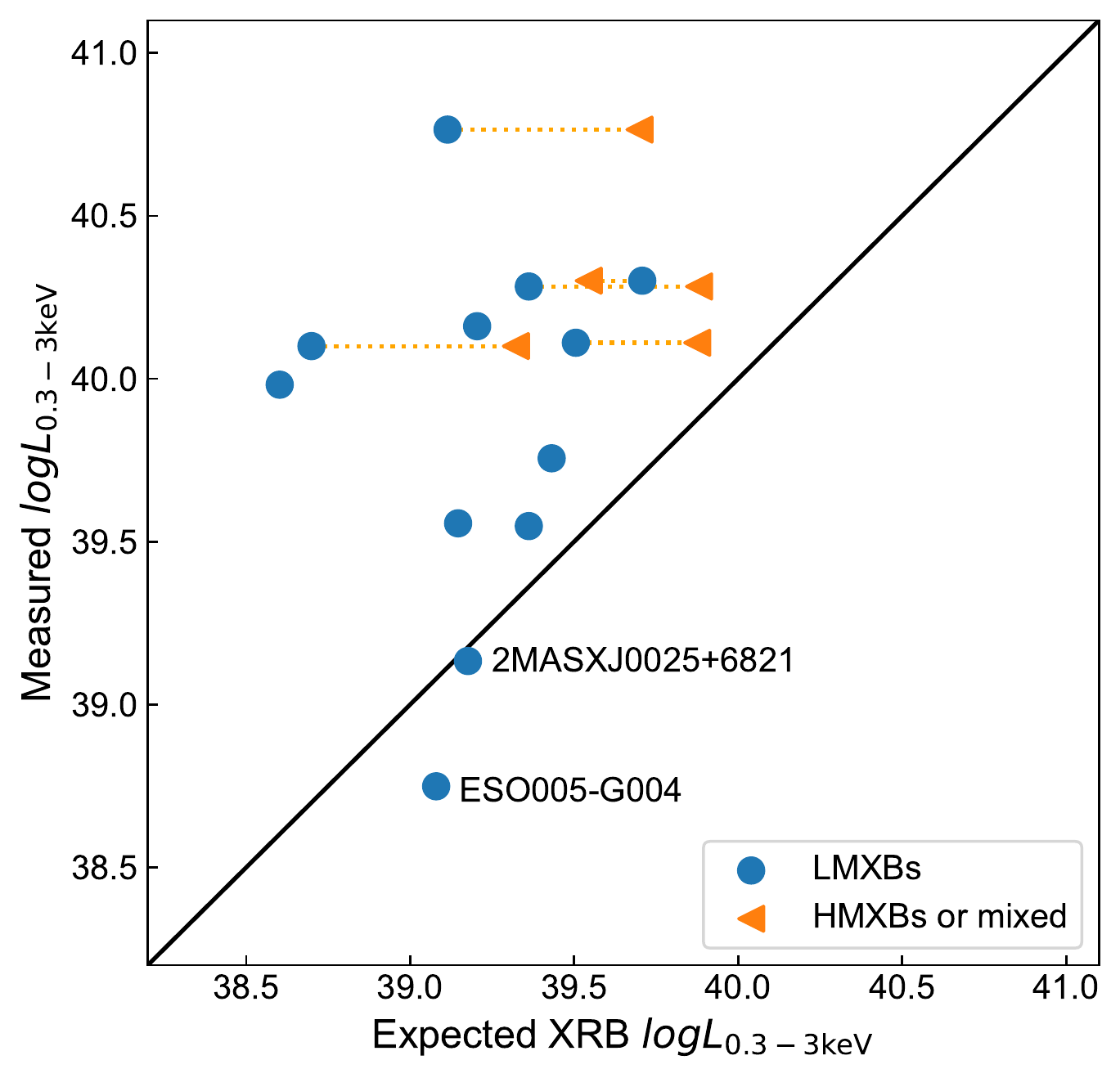}   
\includegraphics[width=8.5cm]{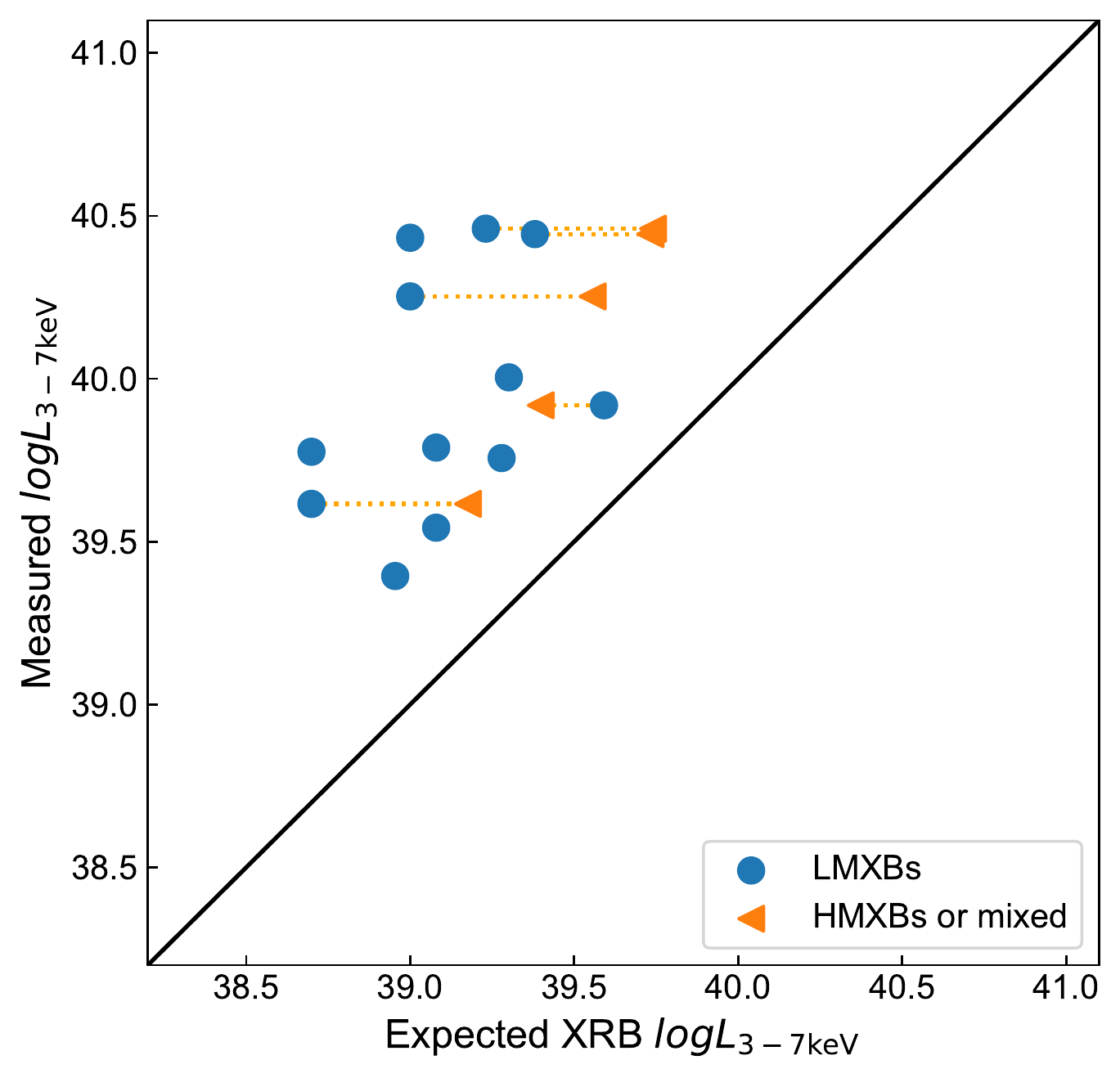}  
\caption{Measured extended X-ray luminosity versus expected luminosity from XRBs for the 0.3-3.0 keV (Left) and 3.0-7.0 keV bands (Right). The blue circles are the expected X-ray luminosities from LMXBs. The orange triangles are the expected X-ray luminosities from HMXBs or mixed populations. The dotted lines show the range between the expected $L_{\rm X}^{\rm LMXB}$ and $L_{\rm X}^{\rm HMXB/mixed}$. The diagonal line denotes the 1:1 ratio. }
\label{LX}
\end{figure*}

Could the extended hard X-ray emission be explained by the X-ray binary (XRB) population, which has a fairly hard spectrum? The expected X-ray luminosities from XRBs were estimated by using scaling laws depending upon the dominant XRB populations. In most cases, the X-ray emission is located within the central bulge of the galaxy, where the low-mass XRBs (LMXBs) dominate. Here we used the $L_{\rm X}$-$L_{\rm K}$ scaling relation for LMXBs from \cite{Boroson2011}. Since the X-ray emission only occupies the central region rather than the entire host galaxy, we also corrected the K-band luminosities for the region where X-ray extent is seen (8$\arcsec$ or 17$\arcsec$ circle defined in Section \ref{methods}), using images from the 2MASS survey \citep{Jarrett2003}. For three sources (NGC 424, NGC 1125, and NGC 3281) where the X-ray emitting region encloses both the bulge and spiral arms where star formation occurs, we also estimated the expected X-ray emission from high-mass XRBs (HMXBs). We used the scaling relation between $L_{\rm X}$ and SFR for HMXBs in \cite{Mineo2012,Mineo2014}. The expected HMXB $L_{\rm X}$ was not corrected for the X-ray emitting region due to complex distribution of star forming regions. These $L_{\rm X}^{\rm HMXB}$ estimates can then be considered as upper limits. Whenever both the SFR and stellar mass are available, we also derived the expected $L_{\rm X}$ from the relation with both SFR and stellar mass \citep{Lehmer2019}.

To facilitate comparison with the measured X-ray luminosities in Table \ref{table4}, we list the expected X-ray luminosities in the 0.3-3.0 keV and 3.0-7.0 keV bands separately in Table \ref{table5}, using a typical soft-to-hard ratio of $\sim$1.3 according to the spectral modeling of XRBs (e.g., \citealt{Boroson2011}). Figure \ref{LX} shows the comparison between the measured extended X-ray luminosities (Table \ref{table4}) and the expected luminosities from XRBs (Table \ref{table5}) for both the soft and hard bands.

In all of the sources, the measured extended hard X-ray luminosities are higher than the expected X-ray luminosities due to XRBs by factors of $\sim$ 2 - 25. The X-ray emission from XRBs could well explain some of the hard extents if not all. There is still excess hard X-ray emission that can not be explained by XRBs, suggesting a connection with the AGN. For the extended soft X-ray emission, the measured luminosities in two sources, ESO 005-G004 and 2MASXJ0025+6821, are consistent with the expected X-ray luminosities from XRBs. For the remaining 10 cases, the excess over the XRB prediction ranges from a factor of $\sim$ 1.5 to 25.

In well-studied CT AGN, which show spectacular ionization bicones and remarkable correspondence between high surface brightness X-ray features and the radio jet and optical line emission, it has been suggested that the extended hard emission is due to scattering in the ISM of photons escaping the nuclear region in the direction of the ionization cone (e.g., ESO 428-G014 \citep{Fabbiano2017,Fabbiano2018a,Fabbiano2018b}; IC 5063 \citep{Travascio2021}; see also \citealt{Jones2020}). The ALMA and SINFONI observations of CO(2-1) and molecular hydrogen ($H_2$) of ESO428-G014 have confirmed that the 3-6 keV continuum and Fe K$\alpha$ emission are due to scattering from dense ISM clouds \citep{Feruglio2020}.

The hard emission seen in the cross-cone may indicate a porous torus or the predicted hot cocoon around the nucleus due to jet-ISM interaction. XRBs could potentially provide an alternative explanation for the extended hard X-ray emission in the cross-cone direction. Previous well-studied CT AGN with deep Chandra observations show that the cross-cones account for about 1/3 of the total excess emission in the hard band \citep{Fabbiano2017,Jones2021}. If we assume the same fraction for this sample (sources in Table \ref{table4} that show clear cone-like structures) and also assume half of the expected XRB emission is located in the cross-cone region, then the estimated fractions of the 3-7 keV cross-cone excess emission that could be explained by the expected XRBs are below 30\% for three of them and up to 70\% for ESO 137-G034. 

Since most of the extended hard emission cannot be explained by XRBs, we further explore the origin associated with the AGN and compare with other extended hard X-ray detected AGN in the following sections. 

\subsection{Comparison with extended hard X-ray detected AGN in the literature}

This sample of heavily obscured, though not CT, AGN was selected under the same criteria as the CT AGN sample in \cite{Ma2020} except for the lower absorbing column densities. We also utilize exactly the same metrics to quantify the amount and extent of the extended component. Now we have a larger sample of obscured AGN with systematically measured excess counts and excess fractions to investigate whether the measured quantities correlate with any physical parameters such as AGN bolometric luminosity $L_{\rm bol}$, black hole mass $M_{\rm BH}$, Eddington ratio $\lambda_{Edd}$, $N_{\rm H}$ etc., which may shed light on the origin of the extended hard X-ray emission.

In \cite{Ma2020}, we found a moderate correlation between the total excess fraction and log$N_{\rm H}$ among the 9 CT AGN (also including ESO 428-G014 and NGC 7212), i.e., CT AGN with a higher log$N_{\rm H}$ tend to have a higher total excess fraction. \cite{Jones2021} presented extended hard X-ray emission of several CT AGN with accumulatively deep Chandra observations that enable studies of trends along the ionization cone direction versus the cross-cone direction. \cite{Jones2021} found a strong correlation in the ionization cone between the total (hard X-ray) excess fraction and log$N_{\rm H}$, while the trend is not observed in the cross-cone region. 

Now that we have a larger sample, we add our new sources to Figure \ref{Nh_excfrac} to revisit this potential correlation. The orange squares are the five obscured AGN in this work. The gray circles are the seven CT AGN plus two individually selected CT AGN (ESO 428-G014 and NGC 7212) in \cite{Ma2020}. The total excess fractions of these AGN are measured over all azimuthal angles. We also include in Figure \ref{Nh_excfrac} the CT AGN from \cite{Jones2021} with the excess fractions measured in the ionization cone direction. Our sample in this work greatly expands the log$N_{\rm H}$ parameter space, however, there is still a gap at log$N_{\rm H}$ $\sim$ 23.5 - 23.8 cm$^{-2}$. The previously observed moderate trend in CT AGN by \cite{Ma2020} or the strong correlation in the ionization cone by \cite{Jones2021} is diluted rather than strengthened by the new sources. We still need more heavily obscured AGN that cover a wider range of column densities to better probe this relation. The dilution of the trend could also suggest that CT AGN and non-CT AGN may have different relations, or there could be a hidden parameter that would separate this sample into subsamples in which the correlation might hold.

No correlations have been found between the total excess fraction and other AGN parameters, i.e., $L_{\rm bol}$, $M_{\rm BH}$, or $\lambda_{Edd}$. We also use the estimated fluxes or luminosities of the extended components instead of the total excess fraction, and we do not see trends either.

In addition, we should point out that the Chandra observations of the sources in our survey (the seven CT AGN in \citealt{Ma2020} and the five obscured AGN in this work) are relatively shallow ($\sim$ 10 - 55 ks) compared to the previously individually selected CT AGN (ESO 428-G014 and the five CT AGN in \citealt{Jones2021}), which have accumulatively deep observations ($\sim$ 100 - 540 ks). Therefore, we should use caution when interpreting results from mixed shallow and deep observations.

Since the seven CT AGN in \cite{Ma2020} and the five obscured AGN in this work were systematically selected and observed under the same survey strategies, here we compare these two samples in terms of the detection rate, spatial extent, and total excess fraction of the extended hard X-ray component. In \cite{Ma2020}, five out of the seven CT AGN show extended emission in the 3.0-7.0 keV band detected at $>$ 3$\sigma$ above the PSF. In this sample, we detected the extended hard X-ray emission in three out of the five AGN. There is no significant difference between the average total excess fractions or fluxes/luminosities of the two samples. However, the CT AGN sample, on average, has an apparently larger hard X-ray extent than this non-CT sample. As discussed in \cite{Ma2020} for CT AGN, the amount of nuclear obscuration may be connected to the dense molecular clouds, i.e., CT clouds, in the host galaxy, which provide the materials needed to scatter the X-ray photons and produce the extended hard X-ray component. The extent of the hard X-ray emission also requires that some of the CT clouds must be coming from much father out in the galaxy than just in the torus. This also explains the observed larger extent in the CT AGN than in the non-CT AGN. Nevertheless we must bear in mind that both samples are still small, and we should revisit this after obtaining large enough samples.

\subsection{Implications for torus modeling}

Although this sample presents a relatively smaller hard X-ray extent than the CT AGN sample, these non-CT but still heavily obscured AGN show hard X-ray emission extending to at least $\sim$250 pc in radius, which is beyond the traditional dusty torus in the AGN unified model (e.g., \citealt{Urry1995,Ramos2017}). Using a simple formula in \cite{Barvainis1987}, we estimated the outer torus radii for the five obscured AGN following \cite{Ma2020}, and they are all within 60 pc. Also, the amounts of extended hard X-ray emission relative to the total emission of these heavily obscured AGN are not negligible. Therefore, we should take this component into account to improve torus modeling. 

All the AGN in this sample have high quality, broad band NuSTAR spectra covering 3-79 keV, which will be presented in a future publication. In principle, the obscuring torus geometry such as the scale height, opening angle, inclination angle, and torus covering factor in clumpy torus models \citep{Nenkova2008,Elitzur2008} can all be extracted from (e.g., NuSTAR) X-ray spectral modeling (e.g., \citealt{Murphy2009,Balokovic2018,Tanimoto2019}). However, the existence of reprocessed emission on $>$100 pc scales is currently not accounted for by any published models. Ignoring this new component can lead to biased estimates of the torus covering factor and/or its average torus column density. It is possible to constrain the opening angle from Chandra imaging of the extended X-ray component, and we know the fraction of total emission that belongs to the extended component, which reduce the uncertainties in the remaining torus model parameters.

\subsection{Implications for AGN feedback}

About two dozen CT AGN with prominent biconical narrow line regions have now been imaged with Chandra \citep{Fabbiano2022}. Almost all of these have the bicone intersecting the host galaxy disk and interacting with the host ISM, including shocked radio jets and fluorescing molecular clouds, leading to complex X-ray emission regions \citep{Fabbiano2022}. They are objects in which we directly see AGN Feedback in action. 

The conical structure in IC 1657, however, emerges perpendicular to the host plane. What would naturally have been assumed to be an AGN ionization bicone in the soft X-rays turns out to be consistent with the prediction from X-ray binary scaling relations, first found by \cite{Fabbiano1989,Fabbiano2006}. The extended soft X-ray emission has an X-ray luminosity of $L_{\rm X}$ $\sim$ 1.3 $\times$ 10$^{40}$ erg s$^{-1}$. The predicted full-band $L_{\rm X}$ is 1.3 $\times$ 10$^{40}$ erg s$^{-1}$ using the stellar mass and star formation rate (SFR) from \cite{Lopez2020} and the X-ray/SFR ratio in \cite{Lehmer2019} with $\sim$40\% of $L_{\rm X}$ in the soft band. In contrast, the hard X-ray emission shows more extended emission in the perpendicular direction, in line with the optical cone. Deeper Chandra data is needed to reveal a potentially more complete hard X-ray morphology.  

\cite{Lopez2020} conducted BPT mapping of IC 1657 using the MUSE IFU data and found the spaxels associated with the ionized cone fall in the region occupied by shock ionization according to the predicated line ratios from theoretical models. The line ratios at the cone region are more compatible with the SF-driven wind rather than AGN-driven wind according to \cite{Sharp2010}. Therefore they conclude that shock ionization produced by a SF-driven outflow seems to be the most likely explanation for the ionized cone. 

The origin of the extended hard X-ray emission along the perpendicular direction is elusive. One possibility is that the hard X-ray emission is still within the galaxy disk. There is then plenty of ISM gas for AGN photoionization, or a radio jet\footnote{\cite{Unger1989} observed IC 1657 with the VLA in a hybrid B-C configuration but do not show a map of the radio emission (unresolved).} to shock against. There also could be molecular clouds for the nuclear photons to interact with, as is the most likely explanation for the extended hard X-ray emission in ESO 428-G014 \citep{Fabbiano2017,Fabbiano2018a}. If instead the hard X-ray emission extends beyond the galaxy disk, as hinted by the current Chandra data, there would be little material to interact with. IC 1657 could then be an extragalactic analog of the Milky Way Fermi Bubble where the central super massive black hole probably released vast amounts of energy that powers high-energy jets in the past \citep{Su2010}. Simulations of jet-ISM interactions (e.g., \citealt{Mukherjee2018}) also describe outflows with hot gas of kT $\gg$ 1 keV that do not cool efficiently and might have a low surface density in the hard band. Again, deep Chandra data is required to confirm the extent and reveal a complete morphology of the hard X-ray emission.

The other sources lack multi-wavelength data that inhibit further interpretation. Nevertheless, this newly discovered, extended hard X-ray component opens up a new window to investigating how the supermassive black hole interacts with and impacts the host galaxy.

\vspace{0.8cm}
\section{Summary and conclusions}
\label{summary}

Following the pilot Chandra survey of nearby CT AGN, in this work we performed a Chandra spatial analysis of five uniformly selected non-CT but still heavily obscured AGN to investigate the extended hard X-ray emission by measuring the excess emission counts, excess fractions, and physical scales. Three of them show extended emission in the 3.0-7.0 keV band detected at $>$ 3$\sigma$ above the Chandra PSF with total excess fractions ranging from $\sim$8\% - 20\%. NGC 678 also exhibits extended emission in the 6.0-7.0 keV band, where the Fe K$\alpha$ line dominates. The extent of the hard emission ranges from at least $\sim$250 pc to 1.1 kpc in radius for the sample. We estimated the fluxes and luminosities of the extended components for the CT and non-CT AGN samples.

We estimated potential contributions from XRBs. In the hard band, XRBs fall short by factors of $\sim$ 2 - 25 in all cases. In 10 cases, XRBs cannot explain the extended soft X-ray luminosity. In two sources, the expected X-ray luminosities due to XRBs could explain the soft extent. XRBs could also provide an alternative explanation to the extended hard emission in cross-cone regions. In most cases, there is still excess hard X-ray emission that cannot be explained by XRBs. Based on deep studies of selected CT AGN, this emission is likely to be connected with the interaction of the AGN photons with molecular clouds.

We have compared these new sources with CT AGN in our previous work and find that CT AGN appear to be more extended in the hard band than the non-CT AGN. We also revisited the tentative correlation found between total excess fraction and log$N_{\rm H}$ from the CT AGN sample by adding the new sources. However, the trend is diluted by the new sources. We would need AGN with a wider coverage of log$N_{\rm H}$ to further test this relation. It could also suggest that CT AGN and non-CT AGN may have intrinsically different relations. 

Similar to CT AGN, the amounts of extended hard X-ray emission relative to the total emission of these non-CT but still heavily obscured AGN are not negligible. Therefore, this newly discovered component must be taken into account in torus modeling with NuSTAR to avoid biased estimates of the torus covering factor and/or its average torus column density. Our Chandra spatial analysis will reduce uncertainties in torus modeling. 

This newly discovered extended hard X-ray component also provides a new opportunity to investigate how the supermassive black hole interacts with and impacts the host galaxy. 

Future Chandra observations of a larger sample of heavily obscured AGN as well as multi-wavelength diagnostics are the way to establish this population and fully uncover the origin of this extended component.

\section*{acknowledgments}

We thank the referee for carefully reviewing our manuscript and providing detailed, constructive comments. We thank the CXC team members who helped us with the Chandra PSF investigation: Diab Jerius, Terry Gaetz, Vinay Kashyap, Tom Aldcroft, and Margarita Karovska. This work makes use of data from the Chandra data archive, and the NASA-IPAC Extragalactic Database (NED). The analysis makes use of CIAO and Sherpa, developed by the Chandra X-ray Center; SAOImage ds9; XSPEC, developed by HEASARC at NASA-GSFC; and the Astrophysics Data System (ADS). This work also uses observations made with the NASA/ESA Hubble Space Telescope, obtained from the data archive at the Space Telescope Science Institute. This work was supported by the Chandra Guest Observer program, grant no. GO1-22106X (PI: M.E.), and by the Chandra Theory grant no. TM9-20004X (PI: M.B.). M.B. acknowledges support from the YCAA Prize Postdoctoral Fellowship. WPM acknowledges support from Chandra Guest Observer grants no. GO8-19099X, GO8-19096X and GO1-22088X. This work was partially performed at the Aspen Center for Physics, which is supported by National Science Foundation grant PHY-1607611.

\begin{appendix}
\section{Uncertainties in the simulated Chandra PSF}

All the excess emission measurements in this paper are based on the simulated Chandra point spread function (PSF) created with the ChaRT optics raytrace tool and the MARX tool that includes detector and aspect effects. The adopted PSF models are the best currently available according to the Chandra X-ray Center (CXC) documentation\footnote{https://cxc.harvard.edu/ciao/PSFs/chart2/caveats.html} and to private communications with the PSF experts at the CXC. There are, however, known issues with the simulated Chandra PSF$^{16}$, which could potentially affect the measured excess counts and significance.

A number of the PSF caveats in the CXC documentation$^{16}$ are qualitative, not quantitative. As a result, the uncertainties in the excess counts these caveats may introduce in the simulated PSF are hard to assess. Here we examine the effects on our results due to the documented quantitative PSF uncertainties.

The Chandra PSF consists of a narrow core where specular reflection dominates and an outer core dominated by scattering that falls off roughly as the square of the angular off-axis angle\footnote{https://cxc.harvard.edu/cal/Hrma/rsrc/Publish/Optics/PSFWings/wing\_analysis\_rev1b.pdf}. We look at the caveats for each part of the PSF in turn.

\section{1. PSF Wings ($>$ 10$\arcsec$)}

For energies $>$ 2 keV, the wings of the simulated PSF underpredict the observed surface brightness profile beyond 10$\arcsec$. The analysis ``Wings of the Chandra PSF"\footnote{https://cxc.harvard.edu/ccw/proceedings/02\_proc/presentations/t\_gaetz/} shows a comparison between the simulated PSF (SAOSAC) and the observed surface brightness profiles for the 2.4-2.6 keV and 6.4-6.6 keV bands, respectively. At 2.4-2.6 keV, the simulated PSF underpredicts the observed surface brightness profile beyond $\sim$20$\arcsec$. At 6.4-6.6 keV, the difference at 10$\arcsec$-20$\arcsec$ is at most a factor 2.

Since no studies specifically in the 3-7 keV band are available, we assume a factor of 2 to test for extended excesses above the PSF. For most of our sources, the extended emission does not extend beyond 8$\arcsec$ in radius thus is not affected. For IC 1657 and ESO 234-G050 which extend to 17$\arcsec$ in radius, we estimated the changes in the excess counts in the 3-7 keV band due to this effect. For IC 1657, assuming a factor of 2 higher in the simulated PSF counts, the total excess counts at 3-7 keV decrease by 7.3\% (4.2 counts) and the S/N reduces from 4.6$\sigma$ to 4.2$\sigma$. For ESO 234-G050, the total excess counts at 3-7 keV decrease by 6.8\% (2.1 counts) and the S/N reduces from 2.9$\sigma$ to 2.7$\sigma$, assuming a factor of 2 higher in the simulated PSF counts.

\section{2. Outer Core ($\sim$1$\arcsec$ - 10$\arcsec$)}

According to the Jerius (2002)\footnote{https://cxc.harvard.edu/cal/Hrma/rsrc/Publish/Optics/PSFCore/ARLac-onaxis.pdf} PSF memo Figure 11 based on HRC data for AR Lac, there are deviations of the model from the data at the $\sim$20\% level, albeit with large error bars.

In particular, the 4 points between 2$\arcsec$ and 11$\arcsec$ scatter either side of unity (Figure 11, right) and are clearly inconsistent with a constant value of 1.2, which is the 20\% enhancement that we added to the PSF to re-assess the significance of the excesses we claim in our objects. The three objects that have 3-7 keV excess counts above 3$\sigma$ in Table \ref{table2} would have reduced significance to 3.2$\sigma$ for NGC 678, 3.3$\sigma$ for IC 1657, and 1.0$\sigma$ thus non-detection for NGC 5899, assuming a factor of 1.2 in the outer core. As a caveat, being an HRC observation, the AR Lac data has no energy information and is strongly weighted to energies $<$ 2 keV.

\section{3. Inner Core ($<$1$\arcsec$) Artifact}

There is also a known asymmetric artifact in a small radial range 0.6$\arcsec$-0.8$\arcsec$ that first appeared around 2002 in the Chandra HRC PSF\footnote{https://cxc.cfa.harvard.edu/ciao/caveats/psf\_artifact.html}, and there is evidence that the feature is also present in ACIS data$^{20}$. We used the CIAO tool {\it make\_psf\_asymmetry\_region} to estimate the level of this asymmetry in the 3-7 keV band, and they all remain $<$ 6\%. This artifact would not affect the asymmetric cone-like structures at larger radii, $>$1$\arcsec$. This is especially clear in the cases of IC 1657 and ESO 234-G050.

\section{4. Aspect Blur}

Another concern is the AspectBlur parameter adopted in the PSF simulation ``to avoid pixelization effects in the pseudo-event files"\footnote{https://cxc.cfa.harvard.edu/ciao/threads/marx/\#blurring}\footnote{https://cxc.cfa.harvard.edu/ciao/why/aspectblur.html}. We have been using the default blur value of 0.07$\arcsec$ in {\it simulate\_psf}. Pixelization effects are not a problem when using sub-pixel binning. A larger blur of 0.25$\arcsec$ has been suggested for ACIS-S observations$^{21}$. As noted$^{21}$, this is ``based on a limited set of simulations."

Just three sources were used for the simulations that led to the AspectBlur = 0.25$\arcsec$ suggestion$^{21, 22}$. Each was believed to be point-like at the time. One source, AR Lac, was well fit with an AspectBlur of 0.07$\arcsec$. AR Lac was observed with the HRC for 20 ks. The HRC oversamples the HRMA PSF. The 0.07$\arcsec$ AspectBlur for AR Lac demonstrates that the HRMA optics are not the cause of any blurring. AR Lac also shows that any blurring due to poor aspect solutions is not universal at least in modest length observations. Long observations ($>$ 50 ks) may be more affected\footnote{https://cxc.harvard.edu/proposer/POG/html/chap5.html\#tth\_sEc5.4.3}.

The other two sources, RS Ophiuchi, and $\tau$ Canis Majoris required a larger AspectBlur $\sim$0.25$\arcsec$. RS Ophiuchi was observed for 91 ks with ACIS-S (ObsID 7457) and $\tau$ Canis Majoris (ObsID4469) was observed for 100 ks with ACIS-I. We checked that the aspect solution did not wander in both observations by dividing them into roughly 20 ks chunks and calculating the offsets of the centroids of each chunk from the first. In no case was the offset larger than 0.08 ACIS pixels, 0.04$\arcsec$, and the mean offset was 0.05 ACIS pixels, 0.025$\arcsec$. Neither observation contains significant blurring due to aspect errors.

However, both RS Ophiuchi, and $\tau$ Canis Majoris were later found to be not good point sources. RS Ophiuchi is a symbiotic star with outbursts in 1985, 2006, and 2021). Other symbiotic stars have been found to be extended with Chandra (e.g., \citealt{Karovska2010}). In fact, RS Oph was found to have variable extended structure following the 2006 outburst \citep{Montez2022}. The third source, $\tau$ Canis Majoris, is an O9 star with a strong wind in the rich young cluster NGC 2362 \citep{Stickland1998}. The X-ray emission has been ascribed to interaction with the surrounding ISM \citep{Chlebowski1989}. Moreover, $\tau$ Canis Majoris is part of a complex system including a bright O star companion $\sim$0.095$\arcsec$ - 0.16$\arcsec$ away \citep{Maiz2020,Stickland1998}. These properties make $\tau$ Canis Majoris a poor prospect for being a point calibration source in Chandra.

In some cases, using a blur of 0.25$\arcsec$ produced a wider PSF than the actual data causing a negative excess (e.g., NGC 5899 in this sample and NGC 424 in \citealt{Ma2020}). This result is unphysical and so clearly an overestimate. A similar result was found for the double AGN NGC 6240 \citep{Fabbiano2020}.

\section{5. An empirical consideration}

We have five AGNs in our sample, all observed near the optical axis, to ensure the best PSF. Given this strategy, the PSF is essentially the same in all cases, for a given energy band. Moreover, as shown by the radial profiles, the depths of all the observations are such that we can trace the surface brightness well at least a factor of 10$^{-4}$ from the central bin. If we had detected extensions in all cases, given the uncertainty in the calibration one could have suspected a spurious wing effect. But we do not. In the case where no extent is claimed, the data indeed follow exactly the PSF profile, within statistics. The same is not true for the `extent' cases although the statistics of the observations are similar. This gives us added confidence that our assumptions on the PSF are valid.

\section{6. Conclusions}

In summary, we do not find supporting evidence for using a larger blur than the default value of 0.07$\arcsec$. In the worst case scenario, where an aspect blur of 0.25$\arcsec$ is used in addition to a factor of 2 higher PSF wings and a factor of 1.2 higher in the outer core, we do not detect hard excess emission above 3$\sigma$ in any cases. We judge this case to be unlikely.

Quantifying the Chandra PSF uncertainties is an incomplete and difficult ongoing effort by CXC and is far beyond the scope of this paper. Even finding appropriate point sources to use as calibrators is hard. For example, tidal disruption events in distant galaxies are surely point-like to Chandra but, so far, have extremely soft spectra with almost all the counts below 0.5 keV \citep{Auchettl2017}. The release of the Chandra Source Catalog, CXC2.1\footnote{https://cxc.cfa.harvard.edu/csc/about2.1.html} will make finding appropriate sources easier.

\end{appendix}

\end{document}